\documentclass[12pt,dvipsnames]{article}

\usepackage{setspace}

\usepackage[utf8]{inputenc}
\usepackage{jheppub}
\usepackage{epsfig}
\usepackage{bbm}
\usepackage{bbding}
\usepackage{amsmath}
\usepackage{amsfonts}
\usepackage{amssymb}
\usepackage{mathtools}
\usepackage{dsfont}

\usepackage{graphicx}
\usepackage{longtable}
\usepackage{pdflscape}
\usepackage{subcaption}
\usepackage{psfrag}
\usepackage{hyperref}
 
\usepackage[capitalise]{cleveref}
\usepackage{tikz}
\usepackage[compat=1.0.0]{tikz-feynman}
\usetikzlibrary{positioning,arrows.meta}
\usepackage{pgfplots}
\usepackage{pgfplotstable}
\usepackage{subfiles}
\usepackage{longtable}

\usepackage{tabularx}
\usepackage{ltablex}
\usepackage{booktabs}
\usepackage[export]{adjustbox}
\usepackage{listings}
\usepackage{multirow} 
\usepackage{cancel,slashed}
\usepackage{bbm}
\usepackage{graphicx}
\usepackage{latexsym}
\usepackage{transparent}
\usepackage{mathtools}
\usepackage{array}
\usepackage{makecell}
\usepackage{graphics,psfrag}
\usepackage{placeins}
\usepackage{nowidow}
\usepackage{listings}
\usepackage[normalem]{ulem}
\usepackage{environ}
\usepackage{enumerate}
\usetikzlibrary{positioning,arrows.meta}
\usetikzlibrary{positioning,trees,decorations.pathmorphing,decorations.markings,decorations.pathreplacing,calc,shapes,shapes.geometric,shapes.symbols,patterns,arrows}
\usetikzlibrary{fadings}
\usetikzlibrary{decorations.shapes}
\usepackage{amsmath}
\usepackage{amssymb}
\usepackage{amsthm}
\usepackage{soul}
\newcommand{\minf}{m_{\varphi}}

\newcommand{\vol}{\mathcal{V}}

\newcommand{\eff}{{\rm eff}}
\newcommand{\infl}{{\rm inf}}

\newcommand{\newshortstack}[1]
{\begingroup\renewcommand{\arraystretch}{1.1}
\ifmmode
\begin{array}{c}#1\end{array}%
\else
\begin{tabular}{c}#1\end{tabular}%
\fi
\endgroup}

\pgfplotsset{
        compat=1.9,
        compat/bar nodes=1.8,
    }

\theoremstyle{definition}

\newcommand{\be}{\begin{equation}}
	\newcommand{\ee}{\end{equation}}
\newcommand{\bea}{\begin{eqnarray}}
	\newcommand{\eea}{\end{eqnarray}}

\newcommand{\kah}{\mathcal{K}}

\newcommand{\alphap}{\alpha^\prime}
\renewcommand{\epsilon}{\varepsilon}

\newcommand{\ben}{\begin{enumerate}}
	\newcommand{\een}{\end{enumerate}}
\newcommand{\bei}{\begin{itemize}}
	\newcommand{\eei}{\end{itemize}}
	
	\makeatletter
\tikzset{
    dot diameter/.store in=\dot@diameter,
    dot diameter=3pt,
    dot spacing/.store in=\dot@spacing,
    dot spacing=10pt,
    dots/.style={
        line width=\dot@diameter,
        line cap=round,
        dash pattern=on 0pt off \dot@spacing
    }
}
\makeatother
	
\tikzset{decorate sep/.style 2 args=
{decorate,decoration={shape backgrounds,shape=circle,shape size=#1,shape sep=#2}}}
\newtheorem*{theorem*}{Theorem}

\newcommand{\mc}{\mathcal}

\newcommand{\coma}{\, , \quad}
\newcommand{\fstop}{\, .}
\def\mpl{M_{\rm Pl}}

\renewcommand{\arraystretch}{1.2} 
\definecolor{NRG}{rgb}{0.172549, 0.627451, 0.172549}

\graphicspath{{Figures/}}

\preprint{DESY 24-163
        
       \hspace{10.88cm}KCL-PH-TH/2024-60}

\title{
Preheating Axions in String Cosmology}
\author[a,b]{Jacob M. Leedom,}
\author[a]{Margherita Putti,}
\author[c]{Nicole Righi,}
\author[a]{$and$ Alexander Westphal}

\affiliation[a]{ Deutsches Elektronen-Synchrotron DESY\\ Notkestr. 85, 22607 Hamburg, Germany}
\affiliation[b]{CEICO, Institute of Physics of the Czech Academy of Sciences\\
Na Slovance 2, 182 00 Prague 8, Czech Republic}
\affiliation[c]{Physics Department, King's College London, Strand, London, WC2R 2LS, U.K.}

\emailAdd{leedom@fzu.cz}
\emailAdd{margherita.putti@desy.de}
\emailAdd{nicole.righi@kcl.ac.uk}
\emailAdd{alexander.westphal@desy.de}

\abstract{Certain inflationary models can feature periods of preheating -- an era preceding reheating during which parametric resonance triggers an exponential production of bosons. This non-perturbative process can have significant impact on the history of our universe, with consequences ranging from altered reheating channels to overproduction of dark radiation to overclosure. In this work, we study parametric resonance production of axions in string models of inflation. We find that the kinetic couplings and moduli-dependent axion masses give rise to generalizations of the  Mathieu equation. We study these generalizations and determine the strength of parametric resonance created by such couplings. We then apply this technology to fibre inflation models in Type IIB orientifold compactifications. We find that heavy axions can be copiously produced and avoidance of overclosure results in constraints on the typical fibre inflation parameter space.}

\begin{document}

\maketitle
\pagestyle{plain}

\newpage

\section{Introduction}
\label{sec:intro}
A period of cosmic inflation remains 
an essential part of the current cosmological paradigm due to its ability to provide compelling solutions to a number of early-universe puzzles. This era is typically modeled via the dynamics of a scalar field dubbed the inflaton --- the universe expands as the inflaton undergoes slow-roll in its scalar potential. Inflation ends once the inflaton begins its descent to the minimum of its potential leading to an epoch of damped oscillations. In simple models, these oscillations cease as the inflaton perturbatively decays into either standard model particles directly or dark sector states that eventually reheat the universe.

This scenario of perturbative reheating changes significantly upon the inclusion of a \textit{preheating} era~\cite{Dolgov:1989us,Traschen:1990sw,Shtanov:1994ce,Kofman_1994,Kofman_1997}. During this period, the inflaton can produce its own quanta and/or spectator\footnote{Spectator here referring to scalar particles that do not appreciably contribute to inflationary background dynamics.} scalar particles via the non-perturbative phenomenon of \textit{parametric resonance}. This mechanism requires scalar particles whose masses vary with the inflaton oscillations. If active, parametric resonance leads to explosive growth of scalar particle modes that are in resonance with the inflaton oscillations. In many cosmological models, this behavior is captured by the Mathieu equation~\cite{Mathieu1868}, whose secular dynamics are described via Floquet theory~\cite{ASENS_1883_2_12__47_0}, or its generalization for expanding spacetimes. 

The population of spectator particles produced during preheating can significantly alter the naive cosmology of an inflationary scenario. If this population becomes non-relativistic soon after its production, it can serve as a component of dark matter. Indeed this mechanism has been used as a production mechanism for axions~\cite{Co:2017mop,Harigaya:2019qnl,Ramazanov:2022kbd,Chatrchyan:2023cmz} and ultralight dark photon dark matter~\cite{Agrawal:2018vin,Co:2018lka,Bastero-Gil:2018uel,Dror:2018pdh,Adshead:2023qiw}. Instead, if this population remains relativistic until today, it will contribute to current dark radiation. Both outcomes can lead to constraints on inflationary models --- if too many non-relativistic particles are produced, parametric resonance can eventually overclose the universe. Conversely, if too much dark radiation is produced, one will run afoul of the current bound on $\Delta N_\eff$.

The dynamics of preheating have been well-studied in the cosmology literature, but it has not yet been as widely explored in string models of inflation. There are several motivating reasons to amend this. First, string theory provides a natural setting to study inflation since one begins from a valid theory of quantum gravity and can therefore address fundamental issues of cosmic inflation directly. Among these are the role of protective symmetries for trans-Planckian field ranges tied to high-scale inflation with a detectable level of primordial direct tensor mode production and the structure and size of inflaton-dark sector cross couplings~\cite{Baumann:2014nda,Cicoli:2023opf}. Furthermore, in contrast to simple QFT models, the masses of all states in a string compactification arise from the vacuum expectation values (vevs) of scalar fields. If some subset of these scalar fields participate in inflation, one expects that many fields have masses that vary as the inflaton oscillates at the end of inflation. A natural question then is to understand the prevalence of parametric resonance in the string landscape. The utility of this is twofold. First, models of string cosmology are known to suffer from issues of overproduction of dark radiation~\cite{Cicoli:2012aq,Conlon:2013isa,Cicoli:2018cgu,Cicoli:2022fzy,Hebecker:2014gka,Angus:2014bia}. As mentioned above, parametric resonance can contribute to $\Delta N_{\eff}$ and thereby exacerbate dark radiation issues for string models. Second, if parametric resonance produces a non-relativistic spectator, avoiding dark matter overproduction provides an additional constraint. In both cases, the resulting condition can be used to constrain the spaces of underlying microscopic parameters controlling the string theory constructions of inflation and spectator sectors. Studies of parametric resonance in the string cosmology literature include the self-production of the inflaton in blow-up inflation~\cite{Barnaby:2009wr} and fibre inflation~\cite{Gu:2018akj}. 

In this work, we will instead be focused on the non-perturbative production of axions during inflation in string cosmology. Axions are believed to be a general feature of string compactifications, an expectation enshrined in the notion of the String Axiverse~\cite{Arvanitaki:2009fg,Cicoli:2012sz,Demirtas:2018akl}. As a generic prediction of string theory, the string axiverse is one of the most important tools to tie string theory to experiments, both through standard model couplings and through secondary primordial gravitational wave production~\cite{DAmico:2021vka, DAmico:2021fhz, Dimastrogiovanni:2023juq} based on the mechanism first established in~\cite{Campbell:1992hc,Anber:2009ua,Cook:2011hg,Senatore:2011sp}. Preheating of axions in string-inspired models where the inflaton $\varphi$ is coupled to the axion $\vartheta$ via a quartic coupling $\lambda\varphi^2\vartheta^2$ to the axions was studied in~\cite{Khan:2021ght}. 
Such a coupling is commonly considered in typical preheating models in cosmology. Naively, one might then consider that the usual treatment of parametric resonance in the literature is sufficient to understand the non-perturbative production of axions in string cosmology. However, in this paper we show that this to be insufficient due to the fact that string theory features non-perturbative effects that have no analogue in EFT models. Namely, the extended objects in string theory, such as Euclidean D$p$-branes, can wrap sub-manifolds of the compactification manifold and rise to axion potentials not normally considered in the field theory context.

To motivate this statement, we note that closed string axions have distinguished properties that set them apart from their field-theoretic analogues. In particular, the perturbative shift symmetry of a closed string axion is preserved to all orders in string perturbation theory. Therefore, axion potentials can only be generated via non-perturbative effects. These can include field-theoretic effects, such as gauge instantons, but also wormholes and purely stringy effects such as worldsheet or Euclidean D-brane instantons. The strength of such stringy effects is controlled by the size of the sub-manifold wrapped by the Euclidean object, which in turn is governed by the vev of a modulus. This gives rise to the general leading-order potential for the axions:
\begin{equation}\label{eq:stringeft}
    \frac{\mc{L}}{\sqrt{-g}}\supset   
    - \sum_a\Lambda_a^4 e^{- q^a_i\tau^i}\cos(q^a_i \theta^i)\,.
\end{equation}
Here the $\{\theta^i\}$ are non-canonically normalized axions while the $\{\tau^i\}$ are the moduli controlling the strength of the non-perturbative effects. The $\{q_i^a\}$ denote the instanton charges, and the $\{\Lambda_a\}$ are overall scales that will depend on the $\tau^i$ and may additionally depend on other moduli via 1-loop determinants. In general there are higher-order instanton effects correcting~\cref{eq:stringeft} which we have omitted. 

In a viable compactification, the terms in~\cref{eq:stringeft} will be supplemented in the total Lagrangian by a potential for the moduli $\{\tau_i\}$ that will ensure that they obtain vevs. This in turn provides a mass for the axions via the potential in~\cref{eq:stringeft}. Therefore, if we identify a subset of the $\{\tau_i\}$ with the fields responsible for inflation, it follows that the masses of at least some axions will vary with time as the inflaton(s) oscillate at the end of inflation. Hence all of the ingredients for parametric resonance can be present in string cosmology. However, the equation of motion that follows from~\cref{eq:stringeft} is not the Mathieu equation, but rather a generalization of the Hill equation~\cite{10.1007/BF02417081} (see also~\cite{Magnus}) called the Whittaker-Hill equation. 
Hence, the standard approximations used to study parametric resonance are not sufficient to completely determine the behavior of preheating from string inflation once considering the expanding background.

In this work, we consider precisely this issue. We will study the non-perturbative production of closed string axions in type IIB orientifold compactifications and apply this in the context of fibre inflation. In~\cref{sec:preheating} we review the mathematics of the Mathieu and Whittaker-Hill equation, as well as their generalizations to expanding cosmologies. We also track the subsequent evolution of the produced particles up to the modern era. In~\cref{sec:string} we apply these equations to inflation in string cosmology. As a case study, in~\cref{sec:FibreInf} we will examine fibre inflation in Large volume Scenarios (LVS) and place bounds on such models so that parametric resonance does not ruin the cosmology by overproducing dark matter. 



\section{The Mathematics of Preheating}
\label{sec:preheating}
In this section, we review the mathematics of parametric resonance and its application to cosmology via preheating. To motivate this study, we start by considering models described by the general Lagrangian 
\begin{equation}
    \begin{aligned}
        \frac{\mc{L}}{\sqrt{-g}} &\supset g^{\mu\nu}K_{i\bar{\jmath}}\partial_\mu T^i \partial_\nu \bar{T}^{\bar\jmath} - V(T^i,\bar T^{\bar\imath})\\
        &= g^{\mu\nu}K_{i\bar{\jmath}}(\partial_\mu\tau^i\partial_\nu\tau^{\bar{\jmath}}+ \partial_\mu\theta^i\partial_\nu\theta^{\bar{\jmath}}) - V(\tau^i,\theta^i)\,,
    \end{aligned}
\label{eq:genlag}
\end{equation}
where the $\{T^i = \tau^i + i\theta^i\}$ are a collection of $i=1\ldots N$ complex scalar fields and the noncanonical kinetic terms depend on functions $K_{i\bar{\jmath}}$ of the $\{\tau^i\}$. In subsequent sections we will consider $\mathcal{N}=1$ supersymmetric theories obtained from Calabi-Yau orientifold compactifications of Type IIB string theory. In this context, the $\{T^i\}$ will be the complexified  K\"{a}hler moduli fields representing the scalar components of their associated chiral supermultiplets and $K_{i\bar{\jmath}}$ the K\"{a}hler metric. Generally, the scalar potential $V$ will have an expansion for the inflaton modulus $\tau_\varphi$ of the form
\begin{equation}
\begin{split}
  V(T^i,\bar T^{\bar\imath}) \simeq & V_{\rm inf}(\tau_\varphi,\langle\tau^i\rangle|i\neq1) + V_{N}(\tau_\varphi,\tau^i-\langle\tau^i\rangle|i\neq1)\\
  &\qquad\qquad\qquad\quad\;\;\,+
    \Delta V_{\text{non-pert}}(\tau_\varphi,\theta,\tau^i,\theta^i|i\neq1)\quad, 
\end{split}
\end{equation}
where $V_{\rm inf}(\tau_\varphi,\langle\tau^i\rangle|i\neq1)$ is the inflation-driving potential of the inflaton K\"ahler modulus $\tau_\varphi$, $V_{N}(\tau_\varphi,\tau^i-\langle\tau^i\rangle|i\neq1)$ the perturbative potential of all the remaining K\"ahler moduli, and $\Delta V_{\text{non-pert}}(\tau_\varphi,\theta,\tau^i,\theta^i|i\neq1)$ denotes the whole instanton-generated non-perturbative contribution which couples the moduli $\tau^i$ to the axions $\theta^j$ which appear in the $T^k\equiv\tau^k+i\theta^k$.

From hereon, $T\equiv T^1 = \tau_\varphi+i \theta$ denotes the complex scalar field containing the inflaton $\tau_\varphi\equiv \tau^1$ and its partner axion $\theta\equiv\theta^1$. As these fields will share a non-canonical $\tau_\varphi$-dependent kinetic term, we will call $\varphi$ and $\vartheta$ the canonically normalized fields corresponding to $\tau_\varphi$ and $\theta$, respectively.
We consider the generic case where, after the end of inflation, the inflaton is oscillating around its minimum, so that the potential can be approximated as 
\begin{equation}
    V_{\rm inf}(\varphi)=\frac12 \minf^2\varphi^2\,.
\end{equation}
The equations of motion 
\begin{equation}
    \ddot{\varphi}(t)+3 H \dot{\varphi}(t)+V^{\prime}(\varphi(t))=0, \quad \text{with}\quad H^2(t)=\frac{1}{3 \mpl^2}\left(\frac{\dot{\varphi}^2(t)}{2}+V(\varphi(t))\right)\,,
\end{equation}
are solved by 
\begin{equation}
    \varphi(t)=\varphi_0+\Delta \varphi\frac{1}{t} \cos(\minf t)\,.
\end{equation}
If there is an additional (pseudo)scalar field $\theta$ coupled to $\varphi$, the inflaton field can decay non-perturbatively through parametric resonance in the time interval between the end of inflation and the beginning of reheating.
The equation of motion for $\theta$ reads 
\begin{equation}
 \frac{\partial\mathcal{L}}{\partial \theta}-\partial_\mu \frac{\partial\mathcal{L}}{ \partial (\partial_\mu \theta)}=0\,.
\end{equation}
If the kinetic terms are diagonal, the equation becomes 
\begin{equation}
    \frac{\partial\mathcal{L}}{\partial\theta}=-\sqrt{-g}\frac{\partial V}{\partial\theta}, \quad \quad \partial_\mu\frac{\partial\mathcal{L}}{\partial (\partial_\mu\theta)}=\partial_\mu (\sqrt{-g}K_{T\bar T}{\theta})\partial^\mu \theta+\sqrt{-g}K_{T\bar T} \partial_\mu\partial^\mu\theta\,.
\end{equation}
Therefore, the equation of motion in a flat expanding universe where $\sqrt{-g}=a^3$, reads
\begin{equation}
K_{T\bar T}\left(\ddot{\theta}+3H\dot{\theta}-\frac{\nabla^2\theta}{a^2}\right)-\partial_\mu K_{T\bar T}\partial^\mu\theta+\frac{\partial V}{\partial\theta}=0\,.
\label{eq:geneom}
\end{equation}
It is convenient to decompose the axion field $\theta$ into a spatially homogeneous background and fluctuations as $\theta(t,\vec{x})=\theta_0(t)+\delta\theta(t,\vec{x})$. 
We can then write the coupled differential equations for the axion by expanding the potential up to linear term in the fluctuations $V(\theta+\delta\theta)=V(\theta)+V'(\theta)\delta\theta+\mathcal{O}(\delta\theta^2)$,
\begin{equation}\label{eq:eom}
\begin{dcases}\ddot{\theta}_0+\left(3 H-\frac{\partial_0K_{T\bar T}}{K_{T\bar T}}\right) \dot{\theta}_0+\frac{1}{K_{T\bar T}}V^{\prime}=0\,, \\
\\
\ddot{\delta \theta}+\left(3 H - \frac{\partial_0 K_{T\bar T}}{K_{T\bar T}}\right)\dot{\delta \theta}-\frac{\nabla^2 \delta \theta}{a^2}+\frac{1}{K_{T\bar T}} V^{\prime \prime} \delta \theta=0\,.
\end{dcases}
\end{equation}
We retain terms only up to linear order in the fluctuations $\delta\theta$ because they are sufficient to capture the quantum process of particle production, as reflected in the growth of the occupation numbers of the quantum fluctuations. We canonically normalize the axion as
\begin{equation}
    \mathcal{L}\supset K_{T\bar T} \partial_\mu \theta\partial^\mu \theta=\frac{1}{2} \partial_\mu \vartheta\partial^\mu \vartheta \,,
\end{equation}
where $\vartheta\equiv \theta \sqrt{2 K_{T\bar T}}$. 
We denote $ (\cdots)'=\frac{\partial }{\partial \theta}(\cdots)=\sqrt{2 K_{T\bar T}}\frac{\partial }{\partial \vartheta}(\cdots). $
By setting the initial homogeneous background field to $\vartheta_0\simeq 0$, we can ignore the background equation of motion and focus on the second line in~\cref{eq:eom}, which now reads
\begin{equation}\label{eq:eomchi}
   \ddot{\delta \vartheta}+\left(3 H - \frac{\partial_0 K_{T\bar T}}{K_{T\bar T}}\right)\dot{\delta \vartheta}-\frac{\nabla^2 \delta \vartheta}{a^2}+\frac{1}{K_{T\bar T}} V^{\prime \prime} \delta \vartheta=0\,.
\end{equation}
The fluctuation field can be expressed in terms of time dependent mode functions for the individual Fourier modes 
\begin{equation}
\label{floquet}
\vartheta(t,\vec{x})=\vartheta_0+\delta\vartheta_k(t,\vec{x})\,,\quad \delta \vartheta_k(t, \vec{x})=\int \frac{d^3 k}{(2 \pi)^3}\left[\hat{a}_k \vartheta_k(t) e^{i \vec{k} \cdot \vec{x}}+\hat{a}_k^{\dagger} \vartheta_k^*(t) e^{-i \vec{k} \cdot \vec{x}}\right]\,.
\end{equation}
Since the absolute direction of momenta is not important when assuming homogeneity and isotropy, we are interested only in the absolute values of $\vec{k}$, and in the following we will use $k\equiv|\vec{k}|$.
\cref{eq:eomchi} finally becomes 
\begin{equation}
\label{eqstrings}
      \ddot{\vartheta}_k+\left(3H-\frac{\partial_0K_{T\bar T}}{K_{T\bar T}}\right)\dot{\vartheta}_k+\left(\frac{k^2}{a^2}+2 \frac{\partial^2 V}{\partial \vartheta^2}\right)\vartheta_k=0\,.
\end{equation}

\subsection{The Mathieu Equation and Preheating}
Let us first consider the simple example of parametric resonance where the kinetic term of the axion is canonical, and the inflaton is coupled to $\vartheta$ via the operator
\begin{equation}
    \begin{aligned}
        V_1(\varphi,\vartheta) &= g\vartheta^2\varphi^2\,.
    \end{aligned}
\label{eq:pot1}
\end{equation}
Neglecting the expansion of the universe (i.e. $H=0 $), the inflaton oscillations are simply given by  $\varphi=\varphi_0+\Delta\varphi\cos(\minf t)$. We can then plug this expression and~\cref{eq:pot1} into~\cref{eqstrings}. The resulting axion fluctuation dynamics can 
be analyzed via application of
\emph{Floquet theory}~\cite{ASENS_1883_2_12__47_0}. In this simple case, the fluctuation equation of motion can be rewritten as the \emph{Mathieu equation }
\begin{equation}
    \vartheta_k''+(A_k-2q \cos(2s))\vartheta_k=0\,,
    \label{Mathieu}
\end{equation}
where $s=\minf t/2$, $q=4g\Delta\varphi/m_{\varphi}^2$, and $A_k=4k^2/\minf^2$. Solutions to this equation strongly depend on the parameters $q$ and $A_k$, and exhibit an exponential instability $\vartheta_k\sim e^{\mu_k s}$ when the modes enter a specific frequency band. The first and strongest of these bands is determined by \mbox{$\Delta k= \frac{\minf}{2}\pm q$}. The $\mu_k$ are called \emph{Floquet exponents}: they determine the growth of the solution, and in the present case they can be found analytically.
These instabilities correspond to exponential growth of the occupation numbers of quantum fluctuations $n_k$, which is computed as
\begin{equation}
    n_k=\frac{\omega_k}{2}\left(\frac{|\dot{\vartheta_k}|^2}{\omega_k^2}+|\vartheta_k|^2\right)-\frac{1}{2}\,.
    \label{nkMat}
\end{equation}
It can be easily shown that the occupation number is exponentially sensitive to the Floquet exponents, as $n_k\propto e^{2\mu_k s}$. When $\mu_k> H$, the solution will have an exponential instability even when the expansion of the universe is taken into account, as we will see below.

\subsection{The Hill Equation and Preheating}

We can now generalize the previous section to more complicated potentials. One such example are the potentials discussed in \cref{sec:string} describing the inflating saxion coupled to the axion with a string theory-inspired potential. In this case, one must use the \emph{Hill equation}~\cite{Magnus,10.1007/BF02417081} 
\begin{equation}
    \vartheta_k''+\left(A_k+qF(t)\right)\vartheta_k=0\,,
    \label{hill}
\end{equation}
with $F(t)$ a periodic function. 
This system can be treated again with Floquet theory: we can write the above differential equation as 
\begin{equation}
 \begin{dcases}
	\vartheta_k'=\delta \pi_k\,,\\
	\delta \pi_k'=-\left(A_k+qF(t)\right)\vartheta_k\,.
 \end{dcases}
\end{equation}
We will not exploit the details of the Floquet theory here, instead we will just outline the steps required to determine the Floquet indices. For further details, we refer the reader to e.g.~\cite{Amin_2014}.

First, we derive the period $T$ of the zero mode, which can be found by energy conservation of a periodic oscillation, and computed numerically; then we solve the system $\partial_t \mathcal{O}(t,t_0)=U(t)\mathcal{O}(t,t_0)$ from $t_0$ to $t_0+T$ to obtain $\mathcal{O}(t_0+T,t_0)$.
We can now diagonalize $\mathcal{O}(t_0+T,t_0)$ to obtain the eigenvalues $o_k^s=\lvert o_k^s\rvert e^{i\theta_k^s}$, where $s=1,2$. 
Explicitly, 
\begin{equation}\label{eq:periods}
	\mathcal{O}\left(t_0+T, t_0\right)=\left(\begin{array}{l}
		\vartheta_k^{(1)}\left(t_0+T\right) \vartheta_k^{(2)}\left(t_0+T\right) \\
		\delta \pi_k^{(1)}\left(t_0+T\right) \delta \pi_k^{(2)}\left(t_0+T\right)
	\end{array}\right),
\end{equation}
with the initial condition $\mathcal{O}(t_0,t_0)=\mathds{1}$.
This is equivalent to solving \cref{hill} for the two sets of initial conditions $\{\vartheta_k^{1}(t_0)=1,\, \delta\dot{\varphi}_k^{1}(t_0)=0\}$ and $\{\vartheta_k^{1}(t_0)=0,\, \delta\dot{\varphi}_k^{1}(t_0)=1\}$ from $t_0$ to $ t_0+T $.
The eigenvalues of \cref{eq:periods} are given by 
\begin{equation}
	o_k^{\pm}=\frac{\vartheta_k^{(1)}+\delta \pi_k^{(2)}}{2} \pm\frac{1}{2} \sqrt{\left(\vartheta_k^{(1)}-\delta \pi_k^{(2)}\right)^2+4 \vartheta_k^{(2)} \delta \pi_k^{(1)}}\,,
\end{equation}
 with all quantities evaluated at $t_0+T$. Since $\mathcal{O}(t_0+T,t_0)=\text{exp}[T\mathcal{M}]$, where $\mathcal{M}$ is a time-independent matrix whose eigenvalues are the Floquet exponents 
\begin{equation}
	\mu_k^\pm=\frac{1}{T}[\ln(\lvert o_k^\pm\rvert )+i\theta_k^\pm]\,\fstop
\end{equation}
We have exponentially growing solutions if 
\begin{equation}
	\text{Re}(\mu_k^\pm)=\frac{1}{T}\ln(\lvert o_k^\pm\rvert )>0\,.
\end{equation}
After finding the solutions, one can construct an adiabatic invariant, which has the interpretation of the comoving occupation number of particles $n_k$ in the mode $k$ in an expanding universe:
\begin{equation}
    n_k=\frac{\omega_k}{2}\left(\frac{|\dot{\Theta_k}|^2}{\omega_k^2}+|\Theta_k|^2\right)-\frac{1}{2}\,,
    \label{nk}
\end{equation}
where $\Theta_k\equiv a^{3/2}\vartheta_k$. 
The total number of created particles will then be given by
\begin{equation}
    n_\vartheta(t)=\frac{1}{(2\pi a)^3}\int d^3k\, n_k(t)=\frac{1}{4\pi^2 a^3}\int dk k^2 n_k(t)\,,
    \label{nchi}
\end{equation}
while the energy density of the produced particles is
\begin{equation}
    \rho_\vartheta=\frac{1}{(2\pi)^3 a^4}\int d^3k\, n_k(t) \omega_k=\frac{1}{2\pi^
    2 a^4}\int dk k^2 n_k(t) \omega_k\,.
    \label{rhochi}
\end{equation}
Once the growth of the  quantities $n_\vartheta$ and $\rho_\vartheta$ will come to halt once parametric resonance ceases, they will start redshifting respectively like matter and radiation.

\subsection{Cosmological Impact of Preheating}

The axions produced via parametric resonance do not necessarily couple to the standard model, and could reside in a hidden sector. 
Such particles can be so light that they remain relativistic until today, hence behaving as dark radiation and contributing to the effective number of relativistic species $N_\eff$. If instead they become non-relativistic during the evolution of the universe, they contribute to the dark matter abundance $\Omega_{\vartheta}$. 

In any case, if they become non-relativistic after reheating (their mass is lower than the reheating scale $m_\vartheta<T_{reh}$), the energy transferred to the  axions is given by the inflaton, which in turn loses its energy and hastens the onset of reheating.

Inflation ends once slow roll conditions are violated, and the inflaton starts oscillating around its minimum, around $H_\infl\sim \minf$. We assume that the decay of the inflaton into SM particles occurs perturbatively through the two-body decay processes, such that we can take the decay rate $\Gamma_\varphi$ to be constant in time. However, reheating does not start immediately, as the perturbative decay of the inflaton to SM particles becomes active only when $3H\sim 2\Gamma_\varphi$. We write the total inflaton decay rate as
\begin{equation}
    \Gamma_\varphi=(c_{hid}+c_{vis})\frac{\pi^2}{48}\frac{\minf^3}{\mpl^2}\equiv (c_{hid}+c_{vis})\Gamma_0\,,
\end{equation}
where $c_{vis}$ and $c_{hid}$ are the coefficients giving the visible and hidden sector decays, respectively. In this part of the work, we shall approximate $\Gamma_\varphi\simeq \Gamma_0$. Therefore,  given $\minf< \mpl$, then $H_\infl\gg \Gamma_\varphi$. 

After inflation ends, the inflaton field starts to oscillate around its minimum and behaves as pressureless matter, so that the Hubble parameter falls off as $H\propto a^{-3/2}= t^{-1}$. When $H$ reaches the order of the inflaton decay rate, it starts  decaying perturbatively and reheating begins. 
We can then write 
\begin{equation}
    \label{eq:Hvsdecay}
    2H_{reh}=3\Gamma_\varphi\,.
\end{equation}
To find the reheating temperature, we can use the relation between the dominant contribution to the energy density and temperature:
\begin{equation}
    \rho_{\varphi}^{reh}=g_{reh}\frac{\pi^2}{30}T_{reh}^4\,.
\end{equation}
Tracing the inflaton energy density from its value at the end of inflation  $\rho_\varphi^{end}\simeq 3H^2_{end}\mpl^2$ all the way to reheating, if we assume that the inflaton only decays perturbatively, then $\rho_\varphi^{reh}\simeq 3H^2_{reh}\mpl^2\simeq \frac{4}{3}\Gamma_0^2\mpl^2$. 
However, before the inflaton starts decaying perturbatively, when $H> \Gamma_0$, there is a period of preheating when parametric resonance takes place.
This has to be taken into account when computing the inflaton energy density at reheating, as some energy has been taken up by the excited axion quanta. 
This can be encapsulated by 
\begin{equation}
	\rho_\varphi^{reh}=\rho_{\varphi}^{max}\left(\frac{a_{max}}{a_{reh}}\right)^3=\left(\rho_\varphi^{end}\left(\frac{a_{end}}{a_{max}}\right)^3 -\rho_\vartheta^{max}\right)\left(\frac{a_{max}}{a_{reh}}\right)^3\,,
\end{equation}
where $\rho_\vartheta^{reh}$ is the axion energy density evaluated at reheating. 
Therefore, simplifying the above expression, 
\begin{equation}
	\rho_\varphi^{reh}=\rho_{\varphi}^{end}\left(\frac{a_{end}}{a_{reh}}\right)^3-\rho_\vartheta^{max}\left(\frac{a_{max}}{a_{reh}}\right)^3= \rho_\varphi^{end}(1-r_{\text{PR}})\left(\frac{a_{end}}{a_{reh}}\right)^3\,,
\end{equation}
where $r_{\text{PR}}\equiv\frac{\rho_\vartheta^{max}}{\rho_\varphi^{end}}\left(\frac{a_{max}^3}{a_{end}^3}\right)$, and
\begin{equation}
	\rho_{reh}=g_{reh}\frac{\pi^2}{30}T_{reh}^4=\rho_\varphi^{end}(1-r_{\text{PR}})\left(\frac{a_{end}}{a_{reh}}\right)^3\frac{4}{3}\Gamma_0^2 \mpl^2\,.
\end{equation}
The scale factor $a_{reh}$ denotes the moment which $H$ reaches $\Gamma_0$. Prior to this moment, the scale factor goes as $a\sim t^{2/3}$ like the inflaton (which we take to be the dominant contribution to the energy budget of the universe). 
Thus $H$ scales as $H\sim a^{-3/2}$. From the value of $H$ at the end of inflation, we can assess its value at the onset of reheating via
\begin{equation}
	H_{reh}=H_\infl\left(\frac{a_{end}}{a_{reh}}\right)^{3/2}\simeq \frac{2}{3}\Gamma_0\,.
\end{equation}
Thus,  $\frac{a_{reh}}{a_{end}}=  \left(\frac{3H_\infl}{2\Gamma_0}\right)^{2/3}$.
The reheating temperature can then be computed via 
\begin{equation}
\begin{split}
	T_{reh}=(\rho_{\varphi}^{reh})^{\frac{1}{4}}\left(\frac{30}{\pi^2 g_{reh}}\right)^{\frac{1}{4}}=&\left(\frac{30}{\pi^2 g_{reh}}\right)^{\frac{1}{4}}\left[\rho_{\varphi}^{end}(1-r_{\text{PR}})\right]^{\frac{1}{4}}\left(\frac{a_{end}}{a_{reh}}\right)^{\frac{3}{4}}\\
 \\
 =&\left(\frac{30}{\pi^2 g_{reh}}\right)^{\frac{1}{4}}\left[\rho_{\varphi}^{end}(1-r_{\text{PR}})\right]^{\frac{1}{4}}\left(\frac{2\Gamma_0}{3H_\infl}\right)^{\frac{1}{2}}\,.
 \end{split}
\end{equation}
Given that we are taking the Hubble rate as though the inflaton is dominating the energy budget of the universe, $H_{reh}$ will not change, as it is determined by the decay rate of the inflaton field. What changes will simply be the temperature of reheating: in the same time between the end of inflation and the beginning of reheating the inflaton field has lost energy to produce axions, and therefore will arrive at the beginning of reheating with a lower density with respect to the case where we do not consider preheating. Therefore, the final reheating temperature will be lower.

\subsubsection{Effective number of relativistic species}
We now assume that the parametric resonance will produce axions that are light enough to remain as dark radiation up to the modern era. In such a scenario, the axions contribute to the number of relativistic degrees of freedom. We now estimate their contribution.

The total radiation energy density after electron-positron annihilation reads 
\begin{equation}
		\rho=\sum_i \rho_i=\frac{\pi^2}{30}\left[\sum_{i=bosons} g_i T_i^4 +\frac{7}{8}\sum_{i=fermions} g_i T_i^4\right]= \frac{\pi^2}{30}g_*(T)T^4\,,
\end{equation}
where $g_*(T)=\sum_{bosons}g_i\left(\frac{T_i}{T}\right)^4+\frac{7}{8}\sum_{fermions}g_i\left(\frac{T_i}{T}\right)^4$. 

In the absence of particle states beyond the standard model, the total radiation energy density comes from photons and neutrinos. Neutrinos remain in thermal equilibrium with the CMB until their interaction rate with other SM particles drops below the expansion rate. After decoupling, the neutrino temperature $T_\nu$ remains approximately equal to the CMB temperature until electron-positron annihilation, which causes the CMB temperature to rise, while it leaves the neutrino temperature nearly unaffected. Assuming instantaneous neutrino decoupling, $ T_\nu/T = (4/11)^{1/3}$.
The radiation energy density can be written as 
\begin{equation}
	\rho_R=\frac{\pi^2}{15}\left[1+\frac{7}{8}N_\nu \left(\frac{T_\nu}{T}\right)^{4}\right]T^4=\left[1+\frac{7}{8}\left(\frac{4}{11}\right)^{4/3} N_\eff\right]\rho_\gamma\,,
\end{equation}
where $N_\nu$ the number of neutrinos species and
\begin{equation}
	N_\eff=\left(\frac{1}{4}\right)^{4/3}N_\nu \left(\frac{T_\nu}{T}\right)^4
\end{equation}
quantifies the effective number of relativistic degrees of freedom that are not the photon. With three active neutrino species, $N_\nu$ is slightly larger than $3$ if one accounts for relic interactions between electrons and neutrinos during the time of electron-positron annihilation. The resulting value is $N_\eff\simeq 3.046$, which also incorporates finite temperature QED corrections to the electromagnetic plasma and flavor oscillations effects~\cite{PhysRevD.56.5123}. 
Various factors constrain the number of effective species. These include the predictions of BBN, paired with observations of light elements
abundances~\cite{Sarkar:1995dd}, CMB temperature and polarization anisotropies~\cite{Lopez:1998aq}, and the large scale structure (LSS) of matter distribution~\cite{Crotty:2004gm}. Within current experimental bounds, all the aforementioned probes show agreement with the standard prediction of \mbox{$N_\eff=3.046$}. On the other hand, current limits allow for deviations from the SM prediction (i.e. for a non-zero $\Delta N_\eff \equiv N_\eff - 3.046$) which would signal new physics. Current bounds constraint $\Delta N_\eff<0.226$~\cite{Yeh:2022heq}. Future observations are expected to greatly improve on the present bounds (see e.g. \cite{CMB-S4:2016ple}).

Taking into account the presence of relativistic axions, the total amount of energy density of the universe reads
\begin{equation}
	\rho_R=\left[1+\frac{7}{8}\left(\frac{4}{11}\right)^{4/3}N_\eff\right]\rho_\gamma+\rho_\vartheta\,.
\end{equation}
By comparing it to
\begin{equation}
	\rho_R=\left[1+\frac{7}{8}\left(\frac{4}{11}\right)^{4/3}(N_\eff+\Delta N_\eff)\right]\,,
\end{equation}
we find 
\begin{equation}
	\Delta N_\eff=\frac{8}{7}\left(\frac{T}{T_\nu}\right)^4\frac{\rho_\vartheta}{\rho_\gamma}\,,
\end{equation}
where $\rho_\gamma$ is the energy density given by photons. We can then write the equation as~\cite{Garcia:2022vwm}
\begin{equation}
	\Delta N_\eff =\frac{120}{7 \pi^2}\left(\frac{11}{4}\right)^{4/3} \frac{\rho_{\vartheta}^{max} }{T^4}\left(\frac{a_{end}}{a_{reh}}\right)^4 \left(\frac{a_{reh}}{a}\right)^4 \left(\frac{a_{max}}{a_{end}}\right)^4 \,.
\end{equation}
After inflation, when $t_{end} < t < t_{reh}$, the inflaton oscillations redshift approximately as pressureless matter, $\rho_\varphi(t)\propto a(t)^{-3}$. Furthermore, we can use adiabaticity in the expansion after reheating, so that $a_{reh}T_{reh}=a T$. Thus,
\begin{equation}
	\Delta N_\eff =\frac{120}{7 \pi^2}\left(\frac{11}{4}\right)^{4 / 3} \left(\frac{\rho_{\varphi}^{reh}}{\rho_{\varphi}^{end}}\right)^{4/3} \left(\frac{T}{T_{reh}}\right)^4\left(\frac{a_{max}}{a_{end}}\right)^4   \frac{\rho_{\vartheta,max }}{T^4}\,.
\end{equation}
The inflaton energy density during reheating is $\rho_{\varphi}^{reh}=g \frac{\pi^2}{30}T_{reh}^4$, which once substituted in the previous equation gives
\begin{equation}
	\label{Dneff}
	\begin{aligned}
		\Delta N_\eff    &=\frac{4}{7}\left[g(T)^4\left(\frac{11}{4}\right)^4 \frac{\pi^2}{30}\left(\frac{\minf^4}{\rho_{\varphi}^{end}}\right)\left(\frac{T_{reh}}{\minf}\right)^4\right]^{1 / 3}\left(\frac{\rho_{\vartheta}^{max }}{\rho_{\varphi}^{end}}\right)\left(\frac{a_{max}}{a_{end}}\right)^4 \,
		.\end{aligned}
\end{equation}
Note that our computation relies on the assumption that we are operating in a regime where the inflaton energy density dominates. 
A natural question arises when particle production reaches such significant levels that the primary contributor to the universe's energy content becomes the generated particles themselves. Yet, as previously discussed, once the energy density of these particles approaches that of the inflaton, effects like backreaction and inflaton fragmentation come into play and can no longer be overlooked~\cite{Kofman_1994,Kofman_1997}. These phenomena act to halt parametric resonance, slowing down the growth of the axion energy density.
Therefore, \cref{Dneff} is a valid approximation for the analysis of this work.

\subsubsection{Dark matter}
Parametric resonance gives rise to relativistic particles during preheating. However, they can be heavy enough that they become non-relativistic during the evolution of the universe, and therefore will not contribute to the effective number of relativistic species $N_\eff$.  
These axions will then constitute some portion of dark matter. 

The relic abundance of dark matter today is given by 
\begin{equation}
	\Omega_\vartheta=\frac{m_\vartheta n_\vartheta(a_0)}{\rho_c}\,,
\end{equation}
where $\rho_c$ is the critical energy density at the present time. We can compute the number density via \cref{nchi}; it grows until $n_\vartheta^{max}$ and then redshifts as $ a^{-3}$. Therefore, assuming entropy conservation after reheating, we find 
\begin{equation}
	\begin{aligned}
		n_\vartheta(a_0)=n_\vartheta^{max}\left(\frac{a_{max}}{a_0} \right)^3=n_\vartheta^{max}\left(\frac{a_{max}}{a_{end}}\right)^3\left(\frac{2\Gamma_\varphi}{3 H_\infl}\right)^{2}
		\left(\frac{T_0}{T_{reh}}\right)^3\,\fstop
	\end{aligned}
\end{equation}
Using $T_0\simeq10^{-31}\mpl$, the number density evaluated today is
\begin{equation}
	n_\vartheta(a_0)\simeq10^{-93}n_{\vartheta}^{max}\left(\frac{a_{max}}{a_{end}}\right)^3\left(\frac{\pi^2}{72}\right)^{2}\left(\frac{\minf}{\mpl}\right)^{4}\left(\frac{\mpl}{T_{reh}}\right)^3\,.
\end{equation}
Hence 
\begin{equation}
\label{omegachi}
	\Omega_\vartheta h^2\simeq 3.8 \times 10^{23} \left(\frac{a_{max}}{a_{end}}\right)^3 \left(\frac{m_\vartheta}{\mpl}\right)\left(\frac{\minf}{\mpl}\right)^{4} \left(\frac{\mpl}{T_{reh}}\right)^3 \left(\frac{n_\vartheta^{max}}{\mpl^3}\right)\,.
\end{equation}
If we take the inflaton mass to be $\minf\simeq 5\times 10^{-5}\mpl$, and taking into account that parametric resonance ends after the end of inflation so that  $a_{max}>a_{end}$, we find a lower bound for the value of the dark matter abundance:  
\begin{equation}
	\Omega_\vartheta h^2\gtrsim  10^{7} \left(\frac{m_\vartheta}{\mpl}\right)\left(\frac{\mpl}{T_{reh}}\right)^3 \left(\frac{n_\vartheta^{max}}{\mpl^3}\right)\,.
\end{equation}
The above is valid only as long as the produced dark matter does not overcome radiation before matter-radiation equality. 
It can indeed happen that the amount of dark matter produced \textit{overcloses} the universe: if the produced particles become non-relativistic soon after or around reheating, their energy density will redshift like $\sim a^{-3}$, whereas the radiation produced by reheating redshifts like $\sim a^{-4}$. At some point, the two will be comparable. If the ratio of the two energy densities $\rho_{\vartheta}/\rho_{r}$ becomes comparable before matter-radiation equality, the evolution of the universe changes and the above estimate is no longer valid. Let us then evaluate when this will happen by taking the following equality, assuming it stays relativistic until $a_{nr}$:
\begin{equation}
	\begin{aligned}
		\rho_{\vartheta}^{reh}\left(\frac{a_{reh}}{a_{nr}}\right)^4\left(\frac{a_{nr}}{a_{eq}}\right)^3&= \rho_{r}^{reh}\left(\frac{a_{reh}}{a_{eq}}\right)^4\\
  \\
	\Rightarrow	\rho_\vartheta^{max}\left(\frac{a_{max}}{a_{end}}\right)^4\left(\frac{a_{end}}{a_{reh}}\right)^4\left(\frac{a_{reh}}{ a_{nr}}\right)^4\left(\frac{a_{nr}}{a_{eq}}\right)^3&=\rho_r^{reh} \left(\frac{a_{reh}}{ a_{eq}}\right)^4\\
  \\
\iff		\rho_\vartheta^{max}\left(\frac{a_{max}}{a_{end}}\right)^4\left(\frac{a_{end}}{a_{reh}}\right)^4&=\rho_r^{reh} \left(\frac{a_{nr}}{ a_{eq}}\right)
		\,.
	\end{aligned}
\end{equation}
Assuming all the remaining energy density of the inflaton goes into the radiation bath, $\rho_\varphi^{reh}=\rho_r^{reh}$, then 
\begin{equation}
\begin{aligned}
	\rho_\vartheta^{max}\left(\frac{a_{max}}{a_{end}}\right)^4\left(\frac{a_{end}}{a_{reh}}\right)^4&=\rho_\varphi^{reh} \left(\frac{a_{nr}}{ a_{reh}}\right)\left(\frac{a_{reh}}{ a_{eq}}\right)\\
 \\
&=\rho_\varphi^{max}\left(\frac{a_{max}}{a_{end}}\right)^3\left(\frac{a_{end}}{a_{reh}}\right)^3\left(\frac{a_{nr}}{ a_{eq}}\right)
	\,.
\end{aligned}
\end{equation}
Therefore 
\begin{equation}
	a_{eq}=\frac{\rho_\varphi^{max}}{\rho_\vartheta^{max}}a_{nr}\frac{a_{end}}{a_{max}}\frac{a_{reh}}{a_{end}}\,.
\end{equation}
This formula encapsulates also the possibility that the produced axions become non-relativistic before the end of reheating by simply taking $a_{nr}=a_{max}$ and $a_{eq}=\frac{\rho_\varphi^{max}}{\rho_\vartheta^{max}}a_{reh}$.
We want to compare this with $a_{eq}^{\Lambda\text{CDM}}=\frac{\rho_{r,0}}{\rho_{m,0}}a_0\simeq 3\times 10^{-4}a_0$. 
Thus,
\begin{equation}
	\label{matdom}
 \begin{aligned}
	\alpha_0\equiv \frac{a_{eq}}{a_{eq}^{\Lambda\text{CDM}}}&\simeq \frac{\rho_\varphi^{max}}{\rho_\vartheta^{max}}\frac{1}{3}\times 10^4 \left(\frac{a_{nr}}{a_0}\right)\frac{a_{end}}{a_{max}}\frac{a_{reh}}{a_{end}}\\
 \\
 &\simeq 3.33\times 10^{3}\,\frac{\rho_\varphi^{max}}{\rho_\vartheta^{max}} \left(\frac{T_0}{T_{nr}}\right) \frac{a_{end}}{a_{max}}\left(\frac{72 \mpl^2}{\pi^2 \minf^2}\right)^{2/3}\,.
 \end{aligned}
\end{equation}
If $\alpha_0\geq1$ the history of the universe is in agreement with $\Lambda$CDM.
If, instead, $\alpha_0\ll1$ the produced axions are overclosing the universe, and changing its history with respect to Big Bang cosmology. There are three different possible scenarios, which we shortly delineate here.

First, if the axions do not interact with other particles and remain stable, they contribute to the dark matter density of the universe. In this case, if their density is too high, it could lead to the overclosure of the universe, thus imposing stringent constraints on the model parameters to avoid such a scenario, as in the case above.
We can use \cref{matdom} to find the maximum value of the axion mass such that we are not obtaining an early matter radiation equality by imposing $a_{eq}<a_{eq}^{\Lambda\text{CDM}}$. 

The second case is when the axions can decay into other massive particles. The corresponding cosmological implications depend on whether these decay products are relativistic or non-relativistic at the time of decay. If the decay products are non-relativistic, they effectively behave like cold dark matter, similar to the stable axion scenario, and the universe may still face the risk of overclosure. On the other hand, if the decay products are relativistic, they will initially redshift as radiation. As they become non-relativistic, they transition to behaving like matter, modifying the redshift dynamics and slightly relaxing the constraints on the model since the energy density redshifts more rapidly when the particles are relativistic.

We can modify \cref{matdom} to bound the axion mass assuming instantaneous decay and assuming that all the energy density in the axion decays into the heavy fields $\rho_\lambda^{dec}=\rho_{\vartheta}^{dec}$. 
The new matter-radiation equality condition, distinguishing the case where the axion decays after reheating (and after becoming non-relativistic) or before reheating (being still relativistic),\footnote{We should also consider the cases where the axion decays  remaining relativistic, but it can be encapsulated in the first case by choosing $a_{nr}=a_{dec}$ and in the second $a_{nr}=a_{reh}$. } reads 
\begin{equation}
\begin{dcases}
    \rho_\vartheta^{max}\!\left(\!\frac{a_{max}}{a_{end}}\!\right)^4\!\left(\!\frac{a_{end}}{a_{reh}}\!\right)^4\!\left(\!\frac{a_{reh}}{a_{nr}}\!\right)^4\!\left(\!\frac{a_{nr}}{a_{dec}}\!\right)^3\!\left(\!\frac{a_{dec}}{a_{NR}}\!\right)^4\!\left(\!\frac{a_{NR}}{a_{eq}}\!\right)^3\!=\rho_r^{reh}\!\left(\!\frac{a_{reh}}{a_{eq}}\!\right)^4\!&\!\!\!\text{if } T_{reh}>T_{dec}\\
    \\
   \rho_\vartheta^{max}\!\left(\!\frac{a_{max}}{a_{end}}\!\right)^4\!\left(\!\frac{a_{end}}{a_{nr}}\!\right)^4\!\left(\!\frac{a_{nr}}{a_{reh}}\!\right)^3\!\left(\!\frac{a_{reh}}{a_{dec}}\!\right)^3\!
   \left(\!\frac{a_{dec}}{a_{NR}}\!\right)^4\!
  \left(\!\frac{a_{NR}}{a_{eq}}\!\right)^3\!=\rho_r^{reh}\!\left(\!\frac{a_{reh}}{a_{eq}}\!\right)^4\!&\!\!\!\text{if } T_{reh}<T_{dec} \,,
\end{dcases}
\end{equation}
where now $a_{dec}$ defines the time at which the axion decays into the heavy bosons, and $a_{NR}$ corresponds to when the latter become non-relativistic.
The two cases are actually the same if we readjust the scale factors. We can then write: 
\begin{equation}
    \begin{aligned}
        \rho_\vartheta^{max}\left(\frac{a_{max}}{a_{end}}\right)^4\left(\frac{a_{end}}{a_{reh}}\right)^4\left(\frac{a_{reh}}{a_{nr}}\right)^4\left(\frac{a_{nr}}{a_{dec}}\right)^3\times \\
        \times \left(\frac{a_{dec}}{a_{NR}}\right)^4\left(\frac{a_{NR}}{a_{eq}}\right)^3&=\rho_\varphi^{max}\left(\frac{a_{max}}{a_{end}}\right)^3\left(\frac{a_{end}}{a_{reh}}\right)^3\left(\frac{a_{reh}}{a_{eq}}\right)^4\\
        \\
    \iff     \rho_\vartheta^{max}\left(\frac{a_{max}}{a_{end}}\right)\left(\frac{a_{end}}{a_{reh}}\right)\left(\frac{1}{a_{nr}}\right)\left(\frac{a_{dec}}{a_{NR}}\right)&=\rho_\varphi^{max}\left(\frac{1}{a_{eq}}\right)\,.
    \end{aligned}
\end{equation}
Therefore we can find the matter-radiation equality scale factor via:
\begin{equation}
\label{aeqnew}
        a_{eq}=\frac{\rho_\varphi^{max}}{\rho_\vartheta^{max}}\left(\frac{a_{end}}{a_{max}}\right)\left(\frac{a_{reh}}{a_{end}}\right)\left(\frac{a_{NR}}{a_{dec}}\right)a_{nr} \,.
\end{equation}
This will place a bound on the mass of the axion, but it will be less constraining with respect to the case in which the axions do not decay at all. 
We can define, similarly to \cref{matdom}, 
\begin{equation}
\label{matdomnew}
    \alpha_1\equiv \frac{a_{eq,1}}{a_{eq}^{\Lambda\text{CDM}}}\simeq \frac{1}{3}\times 10^{4} \frac{\rho_{\varphi}^{max}}{\rho_{\vartheta}^{max}}\frac{a_{end}}{a_{max}}\frac{a_{reh}}{a_{end}} \frac{a_{NR}}{a_{dec}}\frac{a_{nr}}{a_{0}}\,.
\end{equation}
Comparing this with $\alpha_0$ we find
\begin{equation}
    \frac{\alpha_1}{\alpha_0}=\frac{a_{NR}}{a_{dec}}\,.
\end{equation}
The ratio on the right hand side is always greater than one, and so $\alpha_1>\alpha_0$.

The third case is when the decay products are so light that they remain relativistic. They will not contribute to the matter density but will increase the effective number of relativistic species, $\Delta N_\eff$.
The contribution to $\Delta N_\eff$ can be computed as before: 
\begin{equation}
      \Delta N_\eff=\frac{8}{7}\left(\frac{T}{T_\nu}\right)^4\frac{\rho_\lambda}{\rho_\gamma}\,,
\end{equation}
where we identified $\rho_\lambda$ with the energy density of the decay products of the axion. If the decay happens instantaneously, and all energy density is transferred to the latter, $\rho_\lambda^{dec}=\rho_\vartheta^{dec}$. Therefore, 
\begin{equation}
\rho_\lambda^{dec}=\rho_\vartheta^{dec}=\rho_\vartheta^{max}\left(\frac{a_{max}}{a_{end}}\right)^4\left(\frac{a_{end}}{a_{reh}}\right)^4\left(\frac{a_{reh}}{a_{nr}}\right)^4\left(\frac{a_{nr}}{a_{dec}}\right)^3\,,
\end{equation}
such that 
\begin{equation}
\begin{aligned}
\label{dneffdecay}
 \Delta N_\eff&=\frac{120}{7\pi^2}\left(\frac{11}{4}\right)^{4/3}\frac{\rho_\lambda^{dec}}{T^4}\left(\frac{a_{dec}}{a}\right)^4\\
 &\\
 &=\frac{120}{7\pi^2}\left(\frac{11}{4}\right)^{4/3}\frac{\rho_\vartheta^{max}}{T_{reh}^4}\left(\frac{a_{max}}{a_{end}}\right)^4\left(\frac{a_{end}}{a_{reh}}\right)^4\left(\frac{a_{dec}}{a_{nr}}\right)\left(\frac{s_{2}}{s_1}\right)^{1/3}\,.
\end{aligned}
\end{equation}
In the above we used the fact that $ s^{1/3}aT=const$, where $s$ corresponds to the entropy density of the universe, defined as $s\equiv \frac{\rho+p}{T}$. In principle, the entropy density will change if the degrees of freedom present in the universe will change. However, we will not take that into account, and as a first approximation we consider it constant. All these possibilities depend on the axion decay rates through $a_{dec}$, and therefore on the coupling of the axions and on their mass, and in general are very model dependent.

If we compare this with the effective number of relativistic degrees of freedom one would obtain in the case where the axion remains relativistic and never decays $\Delta N_\eff^{rel}$, we have 
\begin{equation}
\label{dnef1}
   \frac{ \Delta N_\eff}{\Delta N_\eff^{rel}}=\frac{a_{dec}}{a_{nr}}\left(\frac{s_2}{s_1}\right)^{1/3}\,.
\end{equation}
%



\section{String axions and preheating}
\label{sec:string}

In this section, we examine parametric resonance in string-inspired cosmological models. Our main focus will be on type IIB O3/O7 orientifold compactifications, which are defined by first compactifying type IIB on a 6-dimensional Calabi-Yau manifold $X_6$ with Hodge numbers $\{h^{1,1},h^{2,1}\}$. To obtain a theory with $\mathcal{N}=1$ SUSY, we must quotient by a combination of worldsheet parity and a holomorphic involution $\sigma$ of $X_6$. The Hodge numbers of the quotient $\widetilde{X}_6\equiv X_6/\sigma$ are split into positive and negative subspaces as $h^{1,1} = h^{1,1}_+ + h^{1,1}_-$ and $h^{2,1} = h^{2,1}_+ + h^{2,1}_-$. For simplicity, we will assume $h^{1,1}_-=h^{2,1}_+=0$. The resulting 4D $\mathcal{N}=1$ EFT will consist of the graviton, axiodilaton, $h^{2,1}_-$ complex structure moduli, $h^{1,1}_+$ K\"{a}hler moduli, and the supersymmetric partners of these fields. Gauge sectors then arise from the inclusion of spacetime filling D3- and D7-branes.

We will assume that fluxes fix the complex structure moduli and axiodilaton at a high scale~\cite{Giddings:2001yu} and so the moduli of the EFT consist of the K\"{a}hler moduli and their axionic superpartners, which form chiral supermultiplets with scalar components $T^j = \tau^j + i\theta^j$. The effective Lagrangian takes the general form of~\cref{eq:genlag} with $N=h^{1,1}_+$ and the kinetic terms determined by the K\"{a}hler metric
\begin{equation}\label{eq_DefKij}
K_{i\bar \jmath}=\partial_{T^i}\partial_{\bar{T}^{\bar\jmath}} K\,,
\end{equation}
where $K$ is the K\"ahler potential. The scalar potential in~\cref{eq:genlag} is then identified as the  F-term scalar potential determined by $K$ and the superpotential $W$ as
\begin{equation}\label{eq_VSUGRA}
    V=e^{K}\left(K^{i\bar{\bar\jmath}}\mathcal{D}_i W\mathcal{D}_{\bar{\jmath}}\overline{W}-3|W|^2\right)\coma
\end{equation}
where $K^{i\bar{j}}$ is the inverse of the K\"ahler metric 
and the K\"ahler covariant derivative is $\mathcal{D}_i W\equiv\partial_i W + K_i W$.

As discussed in the introduction, the axions $\{\theta^i\}$ obtain potentials only via non-perturbative effects. These can enter either the superpotential or the K\"{a}hler potential. In the present work we focus non-perturbative contributions to the superpotential as these are currently better controlled. If $\mathcal{D}$ is a divisor of $\widetilde{X}_6$, then non-perturbative effects can arise from a stack of D7-branes or from a Euclidean D3-brane wrapping $\mathcal{D}$. In either case, the superpotential takes the form
\begin{equation}
    W= W_0 + A_{\mathcal{D}} e^{-a_{\mathcal{D}}T_{\mathcal{D}}}\,,
\end{equation}
where $T_{\mathcal{D}}$ is the complex K\"{a}hler modulus where ${\rm Re}\,T_{\mathcal{D}}$ is the volume of $\mathcal{D}$ and $a_{\mathcal{D}}$ is a constant. $W_0$ and $A_{\mathcal{D}}$ are constants after the axio-dilaton and complex structure moduli are fixed by fluxes.

Inserting this superpotential into~\cref{eq_VSUGRA} one obtains a scalar potential for the axion $\theta_{\mathcal{D}}$ of the form 
\begin{equation}
\label{potentialax}
    V\supset \Lambda_{\mathcal{D}}^4 e^{-a_{\mathcal{D}} \tau_{\mathcal{D}}}\cos\left({a_{\mathcal{D}}\theta_{\mathcal{D}}}\right)\,.
\end{equation}
This potential has the general form displayed in~\cref{eq:stringeft} of the Introduction if we choose $q_{\mathcal D}^i=a_{\mathcal D}\delta_{\mathcal D}^i$. To obtain the usual EFT Lagrangian, we must canonically normalize the axions. This is done by diagonalizing the K\"{a}hler metric and rescaling the axions. Approximating the K\"{a}hler metric to be diagonal with diagonal entry $K_{\mathcal{D}\bar{\mathcal{D}}}$, one has a canonically normalized axion 
\begin{equation}
    \vartheta_{\mathcal{D}} := a_{\mathcal{D}} f_{\mathcal{D}}\, \theta_{\mathcal{D}}\,,
\end{equation}
where 
\begin{equation}
    f_{\mathcal{D}}= \frac{1}{a_{\mathcal{D}}} \sqrt{2K_{\mathcal{D}\bar{\mathcal{D}}}}
\end{equation}
is the decay constant of the axion $\vartheta_{\mathcal{D}}$ such that the axion has periodicity $\vartheta_{\mathcal{D}} \cong \vartheta_{\mathcal{D}} + 2\pi f_{\mathcal{D}}$. If one now takes $\tau_{\mathcal{D}}$ to be the inflaton, then this setup has the necessary ingredients for parametric resonance. 

The setup above realizes the general discussion in the Introduction and will serve as the working example in the subsequent subsections. However, that is not to say that it is the only example of parametric resonance in string compactifications. For example, axions that are not the scalar superpartner of the inflaton may nonetheless obtain masses that vary with time during inflation. This can occur, for example, via kinetic mixing in $K_{i\bar{\jmath}}$ or multi-instanton contributions to the superpotential that furnish the full instanton lattice. There can also be high-order mixing terms in the scalar potential. At larger $h^{1,1}$ parametric resonance arising from such higher-order terms will necessarily be suppressed compared to the typically handful of axions with leading-order couplings to a given inflaton K\"{a}hler modulus furnished by the sparseness structure in the intersection matrices of CY compactifications~\cite{Demirtas:2018akl,Halverson:2019kna}. Furthermore, it is not necessary that the non-perturbative effects must appear only in the superpotential. ED3 instantons may wrap  non-rigid cycles and instead contribute to the K\"{a}hler potential. 

Finally, these considerations can be applied to compactifications of the other perturbative string sectors. For example, in the Type I or Heterotic theories, axions descending from the 10D 2-forms obtain  worldsheet instanton superpotentials from Euclidean F-strings. The strength of such effects are controlled by the volume of 2-cycles given by the vevs of the K\"{a}hler moduli. If the inflaton arises from the K\"{a}hler moduli, then once again parametric resonance can arise.

\subsection{Structure \& dynamics of the Hill equation for string axions}

Building on the preceding discussion, we now focus on the chiral multiplet $T\equiv\tau_\varphi+i\theta$ containing the inflaton $\tau_\varphi$ (vis a vis its canonically normalized pendant $\varphi$) and its partner axion $\theta$ (whose canonical normalization is $\vartheta$). Starting from \cref{potentialax}, after canonical normalization of the axion field, the string theory inspired axion potential can be written as:
\begin{equation}
    V_{ax}=\Lambda^4(\tau_\varphi) e^{-a_\varphi\tau_\varphi}\cos\left(\frac{\vartheta}{f_\vartheta}\right)\,,
\end{equation}
the equation of motion \cref{eqstrings} can be written as:
\begin{equation}
	\ddot{\vartheta}_k+\left(3H-\frac{\partial_0K_{T\bar T}}{K_{T\bar T}}\right)\dot{\vartheta}_k+\left(\frac{k^2}{a^2}+ \frac{\Lambda^4(\tau_\varphi) }{f_\vartheta^2}a_\varphi^2 e^{-a_\varphi\tau_\varphi}\right)\vartheta_k=0\,,
\end{equation}
where $\varphi$ is associated to the inflaton direction. 
In general, at first order, the entry of the K\"ahler metric is $K_{T\bar T}\propto \frac{1}{\langle\tau_\varphi\rangle^2}$ if the inflaton is associated to the ``big" modulus (fibre inflation), while $K_{T\bar T}\propto (\mathcal{V}\sqrt{\tau_\varphi})^{-1}$ if the inflaton is associated to the blow-up modulus (blow-up inflation). 
At first order, the entry of the K\"ahler metric associated to the inflaton field is $K_{T\bar T}\propto\langle\tau_\varphi\rangle^{-2}$. Therefore we can write  $\frac{\partial_0K_{T\bar T}}{K_{T\bar T}}= \beta\frac{\dot{\tau}_\varphi}{\tau_\varphi}$, where $\beta=2$ for fibre inflation.\footnote{We keep the generic factor implicit so that the analysis can be carried over to e.g. blow-up inflation by using $\beta=\frac{1}{2}$.} Finally, the equations of motion become 
\begin{equation}
	\ddot{\vartheta}_k+\left(3H+\frac{\beta \dot{\tau_\varphi}}{\tau_\varphi}\right)\dot{\vartheta}_k+\left(\frac{k^2}{a^2}+ m_{\vartheta,\eff}(\tau_\varphi)\right)\vartheta_k=0\,,
\end{equation}
where
\begin{equation}
    m_{\vartheta,\eff}^2(\tau_\varphi)=\frac{\Lambda^4(\tau_\varphi)}{f_\vartheta^2}  e^{-a_\varphi\tau_\varphi}\,.
\end{equation}
Parametrizing the inflaton oscillations as $\varphi(t)\simeq \langle\varphi\rangle+\Delta\varphi\frac{1}{t}\cos(\minf t)$, and defining $\tilde{\tau}_\varphi\equiv a_\varphi\tau_\varphi$, the mass of the axion field becomes 
\begin{equation}
\label{meff}
m_{\vartheta,\eff}^2(\tau_\varphi)=m_\vartheta^2
 e^{-\Delta\tilde{\tau}_\varphi\frac{1}{s}\cos(2s)}\,,
\end{equation}
where $m_\vartheta^2=\frac{\Lambda^4(\tau_\varphi)}{f_\vartheta^2} e^{-\langle\tilde{\tau}_\varphi\rangle}$, and  $s=\frac{\minf t}{2}$ is a new time variable that counts the number of oscillations of the inflaton field.
We shift the field as
\begin{equation}
\Theta_k\equiv \vartheta_k\frac{t_0}{t}\,,
\end{equation}
where $t_0=\frac{\pi}{4m}$ corresponds to a quarter of oscillation of the inflaton field.\footnote{We start our analysis from $t_0$ in this section in order to trust the approximation of the oscillatory behavior of the inflaton field. Once a specific inflationary model is chosen, one should consider the whole trajectory of the inflaton field starting from the end of inflation, when the slow roll conditions are broken.}
Changing the time variable to $s$,
the equations of motion become 
\begin{equation}
\label{stringtoteom}
	\Theta_k''+\beta \frac{\tilde{\tau}_\varphi'}{\tilde{\tau}_\varphi} \Theta_k'+\left(\frac{4k^2}{\minf^2}\left(\frac{1}{a}\right)^{2}-\beta\frac{\tilde{\tau}_\varphi'}{\tilde{\tau}_\varphi}\frac{\dot{a}}{a}+\frac{4m_{\vartheta,\eff}^2}{\minf^2} 
	\right)\Theta_k=0\,,
\end{equation}
with 
\begin{equation}
    \frac{\tilde{\tau}_\varphi'}{\tilde{\tau}_\varphi}=-\frac{s_0}{s}\frac{\Delta \tilde{\tau}_\varphi(\cos(2s)/s+2\sin(2s))}{\langle \tilde{\tau}_\varphi\rangle+\Delta \tilde{\tau}_\varphi\cos(2s)}\,.
\end{equation}
The efficiency of parametric resonance is governed by the effective mass parameter, which is directly tied to the mass of the axion field. For extremely light axions, the oscillating term is suppressed by the small axion mass, rendering parametric resonance less effective. To contribute to the effective number of relativistic species today, axions must remain in the form of radiation, requiring their mass to satisfy $m_\vartheta \lesssim \text{eV} \simeq 10^{-27} \mpl$. Consequently, the effective mass squared, as defined in \cref{meff}, is suppressed by approximately $\mathcal{O}(10^{-54})$.

This suppression arises from the exponential dependence on the vev of the inflaton field, $e^{-\langle\tilde{\tau}_\varphi\rangle}$. Initially, this suppression can be offset by the wide oscillations of the inflaton field. However, as the oscillations are damped and their amplitude decreases as $t^{-1}$, this compensating effect diminishes rapidly.

In an expanding universe, additional terms from the derivative of the Kähler metric become significant when performing a full analysis. As the inflaton oscillates with a decaying amplitude, the kinetic mixing term in the equation of motion \cref{stringtoteom} initially dominates but weakens over time. When the axion is heavy, such that 
$ \frac{4m_{\vartheta,\eff}^2}{\minf^2} \gtrsim \beta \frac{\tilde{\tau}_\varphi'}{\tilde{\tau}_\varphi}H$, 
the kinetic mixing term introduces only minor oscillations. Conversely, for a light axion with a small $m_{\vartheta,\eff}^2$, this term grows in relative importance, eventually surpassing the mass term. In this regime, the kinetic mixing term becomes the primary driver of parametric resonance. 

Thus, there exists a threshold below which the resonance effect becomes independent of the axion mass. For compactifications that result in extremely small axion masses --- allowing axions to remain as radiation until today --- parametric resonance can still occur even when the effective mass is negligible. However, the overall impact of this resonance in such cases is minor and not observationally significant.

\subsubsection{Mathieu limit}
Let us first ignore the expansion of the universe and set $H=0$, $a=1$. 
In order to have production of particles in the expanding universe, it is indeed a necessary, but not sufficient, condition that the exponential instability be present in the absence of the expansion. 
\begin{equation}
	\vartheta_k''+\beta \frac{\tilde{\tau}_\varphi'}{\tilde{\tau}_\varphi}\vartheta_k'+\left(\frac{4k^2}{\minf^2}+\frac{4}{\minf^2} \frac{\Lambda^4(\tau_\varphi) }{f_\vartheta^2} 
 e^{-\langle \tilde{\tau}_\varphi\rangle }e^{-\Delta \tilde{\tau}_\varphi\cos(2s)}\right)\vartheta_k=0\,.
	\label{eq:no_exp_eom1}
\end{equation}
The damping term $\frac{\tilde{\tau}_\varphi'}{\tilde{\tau}_\varphi}\vartheta_k'=-2\frac{\Delta \tilde{\tau}_\varphi \sin(2s)}{\langle \tilde{\tau}_\varphi\rangle +\Delta \tilde{\tau}_\varphi\cos(2s)}$ at first order in $\Delta \tilde{\tau}_\varphi/\langle \tilde{\tau}_\varphi\rangle $, is a function oscillating between positive and negative values of $\frac{\Delta \tilde{\tau}_\varphi}{\langle \tilde{\tau}_\varphi\rangle}<1$.  

When the displacement is small, $\Delta \tilde{\tau}_\varphi<1$, the equation of motion can be reduced to the Mathieu equation by truncating the exponential series and neglecting the damping term. Therefore, we can write \cref{eq:no_exp_eom1} as
\begin{equation}
	\vartheta_k''+(A_k-2q \cos(2s))\vartheta_k=0\,,
\end{equation}
with $A_k=\frac{4}{\minf^2}(k^2+\frac{\Lambda^4(\tau_\varphi)}{f_\vartheta^2} e^{-\langle \tilde{\tau}_\varphi\rangle } )$ and $q= \frac{2}{\minf^2} \frac{\Lambda^4(\tau_\varphi)}{f_\vartheta^2}  \Delta \tilde{\tau}_\varphi e^{-\langle \tilde{\tau}_\varphi\rangle }$.
Since we are in the regime where $\Delta \tilde{\tau}_\varphi<1$, then the resonance parameter $q$ has to satisfy 
\begin{equation}
	q< \frac{2}{\minf^2}\frac{\Lambda^4(\tau_\varphi)}{f_\vartheta^2} e^{-\langle \tilde{\tau}_\varphi\rangle}\,.
\end{equation}
As we mentioned before, an important feature of the solutions to the Mathieu equation is the existence of an exponential instability $\vartheta_k\propto e^{\mu_k s}$ within the set of resonance bands of frequency $\Delta k$. This instability corresponds to an exponential growth of the occupation numbers of quantum fluctuations $n_k(t)\propto e^{2\mu_k s}$ that can be interpreted as particle production. 
The parameter $q$ determines the amount of parametric resonance, and, in agreement with the Mathieu analysis, when there is narrow resonance ($q<1$) we find the most amount of growth for modes with $k\simeq k_{max}$, where $k_{max}$ is the mode for which $A_k=1$ (this is the so-called stability band), for which $\mu_{k,max}=\frac{q}{2}$. However, given that during the expansion of the universe the momenta will redshift, the narrow bands that characterize the resonance in the $q<1$ regime, will obstacle the production of particles. 
In the limit $q\to 0$ instead there will be no resonance. 

\begin{figure}
	\centering
	\includegraphics[scale=0.61]{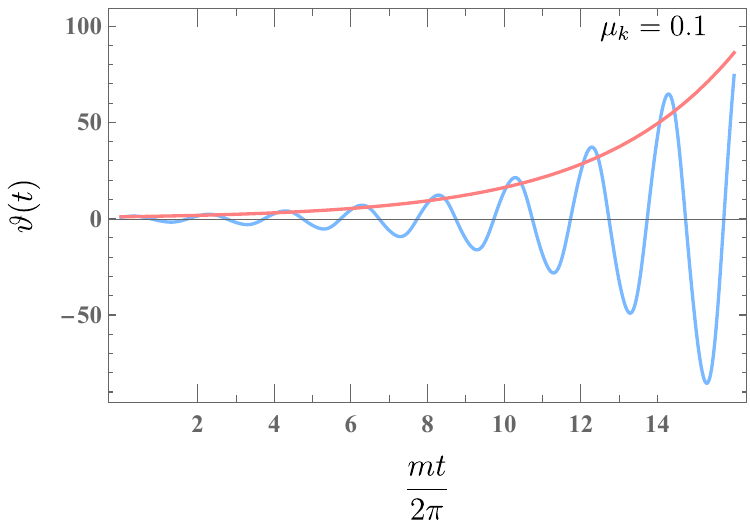}
	\includegraphics[scale=0.6]{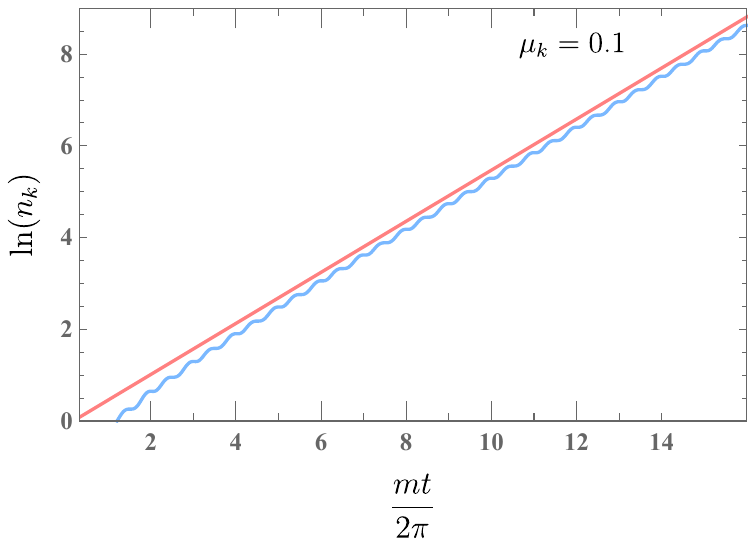}
	\caption{On the left: narrow parametric resonance for the field $\vartheta$ in Minkowski space, for $A_k=1$, and $q\simeq0.19$, for the momentum $k_{max}$ that corresponds to the maximal speed of growth. The pink line corresponds to the solution $\vartheta_k\simeq \exp{(\frac{\mu_k}{2m} t)}$, where $\mu_k=q/2$. On the right: the logarithm of the occupation number of particles $n_k$ in this mode, see \cref{nk}. The number of particles grows exponentially, and the $\ln n_k$ looks like a straight line with a constant slope. This slope, divided by $2\pi$, gives the Floquet exponent $\mu_k$, which in this case is $\mu_k\simeq0.1$. The set of parameters used for these figures are $\Delta \tilde{\tau}_\varphi\simeq \frac{1}{2}$, $\tilde{\tau}_\varphi\simeq 5\pi$, $\frac{\Lambda^4(\tau_\varphi)}{f_\vartheta^2}\simeq 10^{-4}$.}
	\label{fig:1Matthieu}
\end{figure}

\cref{fig:1Matthieu} shows an example of a solution and number density of particles \cref{nk} for this limit, where  $\minf\simeq5\times 10^{-5}\mpl$, and the mass of the axion in this case reads $m_\vartheta\simeq 2.3 \times 10^{-10}\mpl$. 
Lower masses heavily suppress the parameter $q$, and the displacement is not able to offset this suppression. 
Therefore, in the limit of extremely small displacement and narrow resonance, the production of quanta that will stay radiation until now is extremely suppressed.

\subsubsection{Hill limit}

When the displacement of the inflaton field is $\Delta \tilde{\tau}_\varphi>1$, the full equation of motion \cref{stringtoteom} is needed in order to fully capture the complexity of the system. 
We can write the equation at first order in $\frac{\Delta \tilde{\tau}_\varphi}{\langle\tilde{\tau}_\varphi\rangle}$ as
\begin{equation}
	\vartheta_k''-2\beta\frac{\Delta \tilde{\tau}_\varphi}{\langle \tilde{\tau}_\varphi\rangle}\vartheta_k' \sin(2s)+\left(\frac{4k^2}{\minf^2}+\frac{4}{\minf^2} \frac{\Lambda^4(\tau_\varphi)}{f_\vartheta^2} e^{-\langle \tilde{\tau}_\varphi\rangle }
 e^{-\Delta \tilde{\tau}_\varphi\cos(2s)}\right)\vartheta_k=0\,.
	\label{no_exp_Hill}
\end{equation}
\begin{figure}[t]
	\centering
	\includegraphics[scale=0.62]{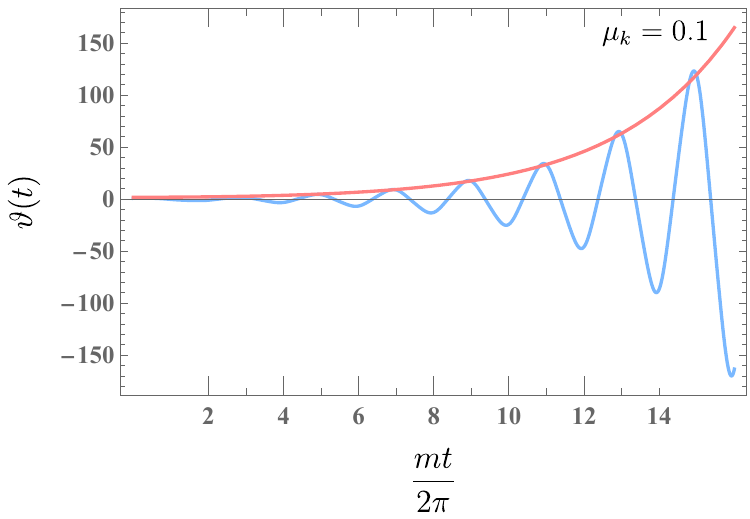}
	\includegraphics[scale=0.59]{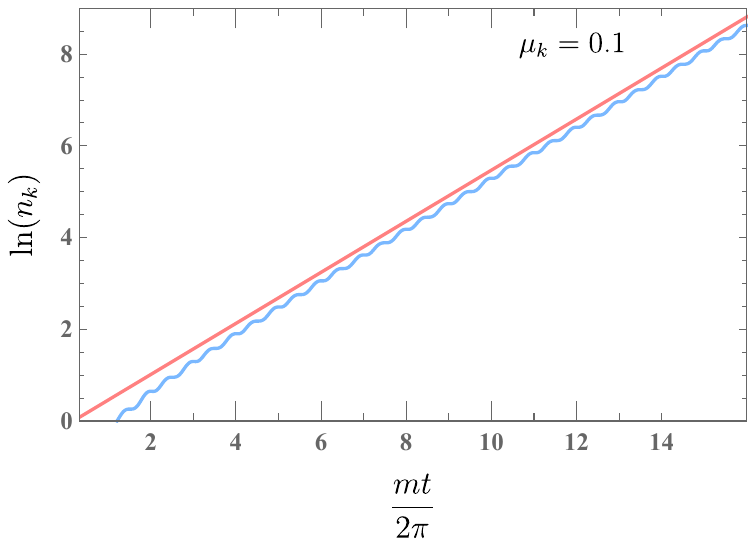}
	\caption{On the left: parametric resonance for the field $\vartheta$ in Minkowski space, for $B_k=1$, for the momentum $k_{max}$ that corresponds to the maximal speed of growth. The pink line corresponds to the solution $\vartheta_k\simeq e^{\frac{\mu_k}{2m} t}$, where $\mu_k$ is the Floquet exponent obtained via the Floquet theorem.   Time is shown as $\frac{s}{\pi}=\frac{mt}{2\pi}$, which is the number of oscillations of the inflaton field. On the right: the logarithm of the occupation number of particles $n_k$ in this mode, see \cref{nk}. The number of particles grows exponentially, and the $\ln n_k$ looks like a straight line with a constant slope. This slope, divided by $2\pi$, gives the Floquet exponent $\mu_k$, which in this case is $\mu_k\simeq0.106$.
 The set of parameters used are $\Delta \tilde{\tau}_\varphi\simeq 6\pi$ and $\langle \tilde{\tau}_\varphi\rangle= 41\pi$, and $m_\vartheta^2=\frac{\Lambda^4(\tau_\varphi)}{f_\vartheta^2}e^{-\langle\tilde{\tau}_\varphi\rangle}\simeq 2\times 10^{-62}\mpl^2$. }
	\label{fig:StringHill}
\end{figure}
In the regime where the amplitude of the inflaton oscillations is small, this equation resembles the Hill equation, and is known as the \emph{Whittaker-Hill equation}: 
\begin{equation}
	\vartheta_k''+2p\sin{(2s)}\vartheta_k'+\left[B_k+2 q F(s)\right]\vartheta_k=0\,,
\end{equation}
with, at first order in $\frac{\Delta \tilde{\tau}_\varphi}{\langle\tilde{\tau}_\varphi\rangle}$, 
\begin{equation}
	\begin{aligned}
		&p=-\beta\frac{\Delta \tilde{\tau}_\varphi}{\langle \tilde{\tau}_\varphi\rangle}\,,\,\,\quad B_k=\frac{4k^2}{\minf^2}\,,\,\,\quad q=\frac{2}{\minf^2} \frac{\Lambda^4(\tau_\varphi)}{f_\vartheta^2}  e^{-\langle \tilde{\tau}_\varphi\rangle }\,,\,\, \quad F(s)=
  e^{-\Delta \tilde{\tau}_\varphi\cos(2s)}\,.
	\end{aligned}
\end{equation}
\begin{figure}[t]
	\centering
	\includegraphics[scale=0.7]{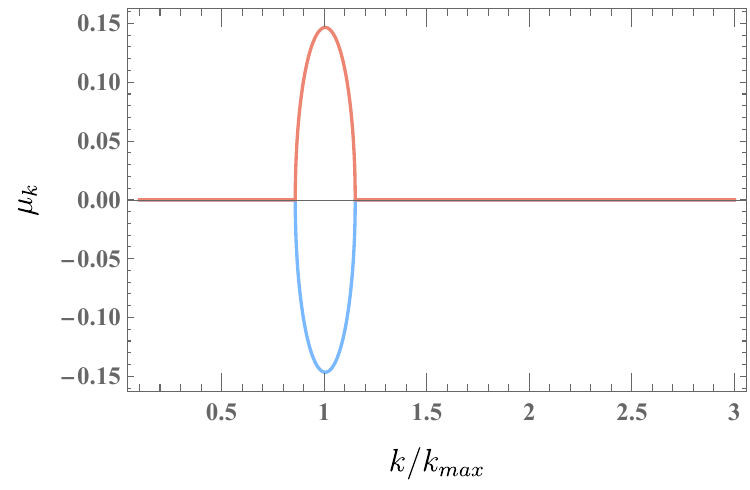}
	\caption{Floquet exponents as a function of the momentum $k$. In this case the exponential instability is only active in one band around the maximum $k_{max}$, for which $B(k)\simeq1$.}
	\label{fig:stab}
\end{figure}
By defining the new function $f(s)=e^{-p \cos(2s)/2}\vartheta_k$, we can write the Whittaker-Hill equation as 
\begin{equation}
	f^{\prime \prime}+\left[B_k-\frac{p^2}{2}-2p \cos (2 s)+\frac{p^2}{2} \cos (4 s)+2q F(s)\right] f=0\,.
	\label{whithil}
\end{equation}
This equation is a second order differential equation with periodic coefficients, and therefore it has the form of the Hill equation. Therefore  
we can study it with a Floquet analysis. The solutions of this equation are 
\begin{equation}
	\vartheta_k(t)=\vartheta_{k_+}(t)e^{\mu_k t}+\vartheta_{k_-}(t)e^{-\mu_k t}\,,
\end{equation}
where $\vartheta_{k_\pm}$ are periodic functions in time and $\mu_k$ are complex coefficients. 
A necessary, but not sufficient, condition for parametric resonance is Re${[\mu_k]}>0$: larger values of Re$[\mu_k]$ indicate stronger parametric resonance. 

\cref{fig:StringHill} shows an example of solutions in this limit for $\minf\simeq 5\times 10^{-5}\mpl$ and a very light produced axion: $m_\vartheta\sim 1.5\times 10^{-31}\mpl$. The axion therefore stays relativistic up to today, and contributes to $\Delta N_\eff$.  

\cref{fig:stab} shows the momenta that get excited and their corresponding Floquet exponent. We can see that the peak of the excited momenta is around $k\sim \frac{m_\varphi}{2}$, as expected in the Mathieu case. However, in contrast to the Mathieu case, the other momenta in the secondary bands are much less excited and therefore their contribution will be negligible. 
\begin{figure}[t!]
	\centering
	\includegraphics[scale=0.54]{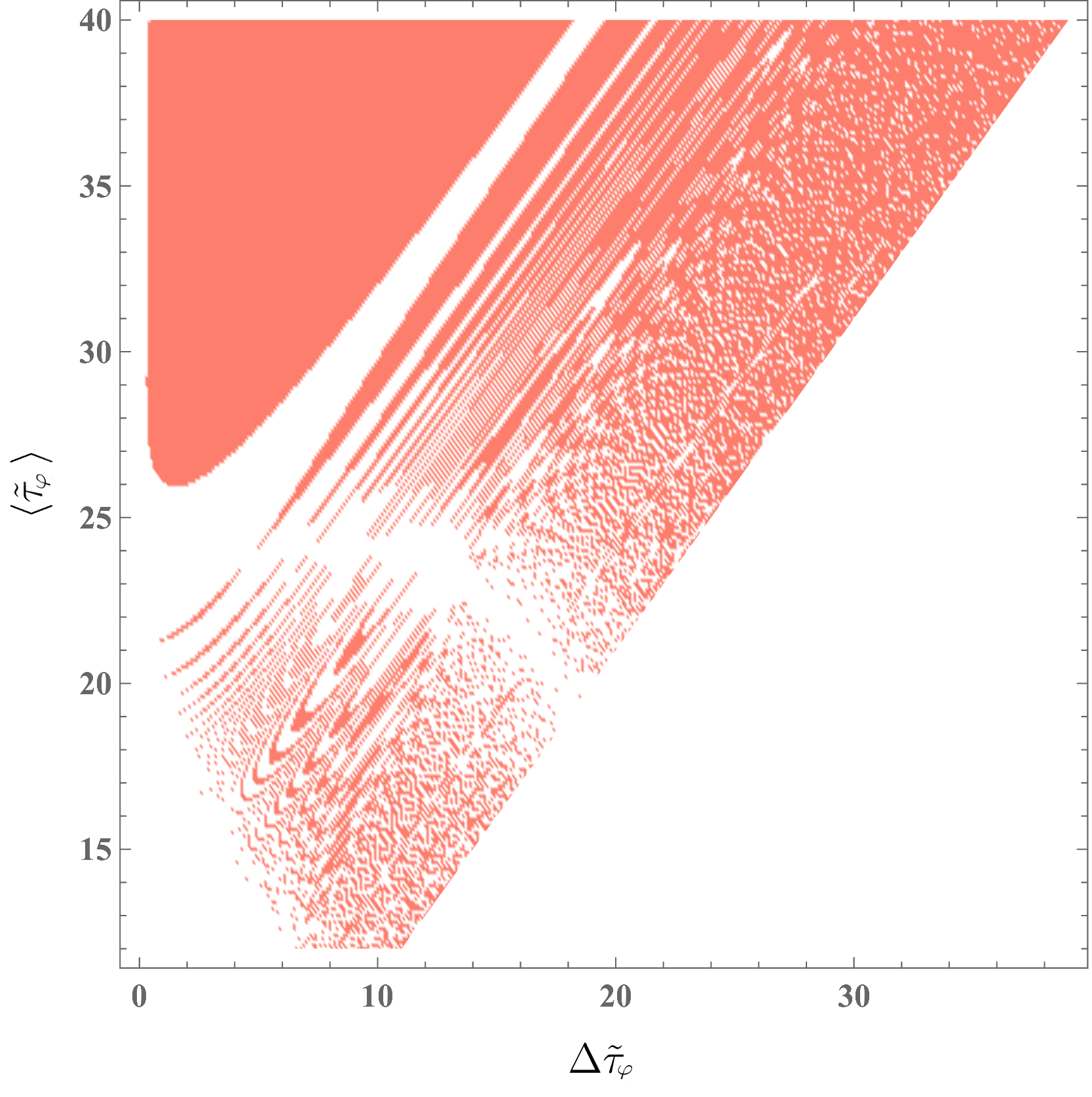}  
 \hfill
 \includegraphics[scale=0.54]{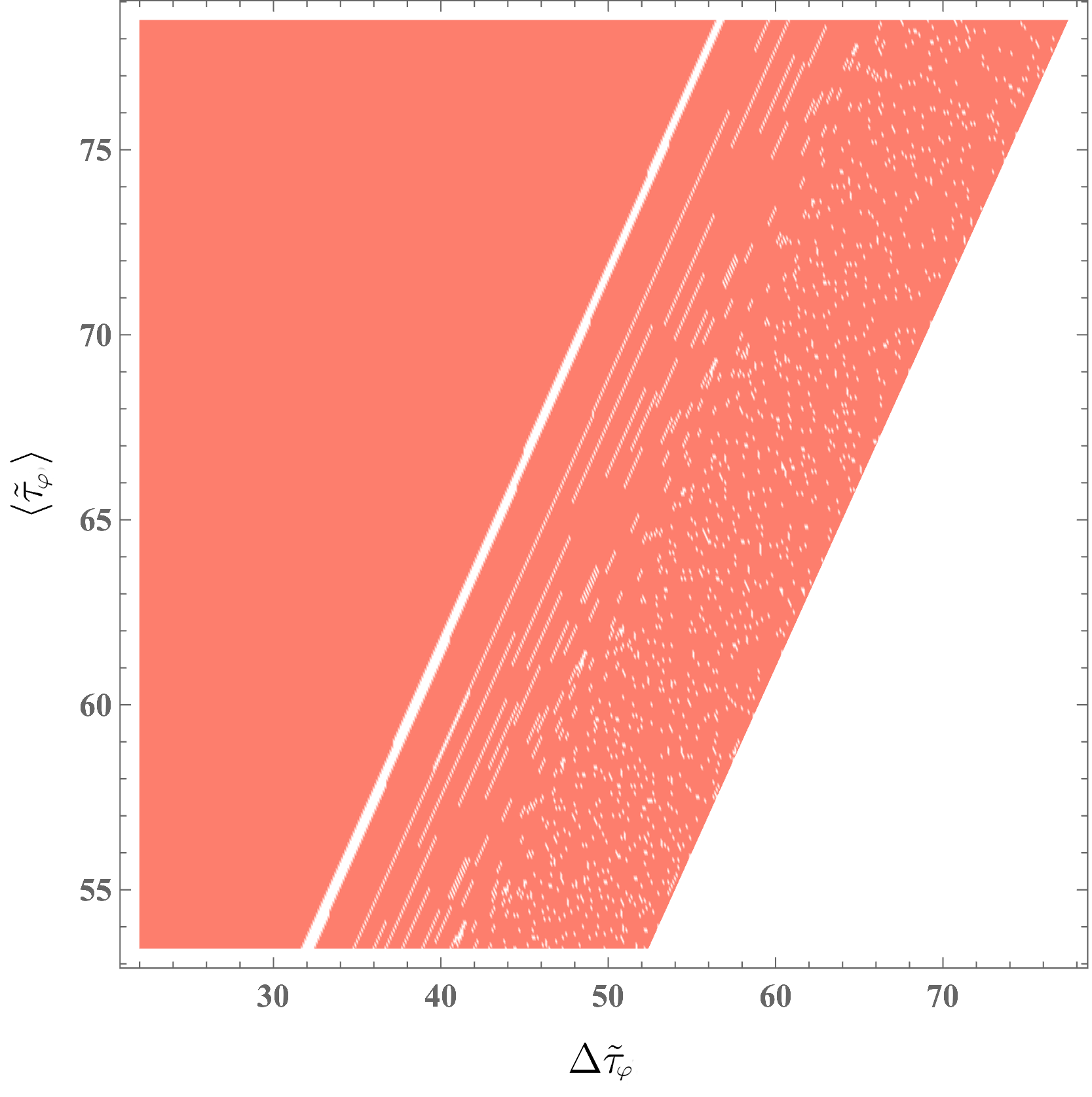}
	\caption{Stability charts for the Hill equation at $k=k_{max}$: the plot shows in red the points of $\tilde{\tau}_\varphi,\,\Delta \tilde{\tau}_\varphi$ for which the Floquet exponent $\mu_k\geq 0.01$ for $\frac{\Lambda^4(\tau_\varphi)}{f_\vartheta^2}\simeq 4\times 10^{-6}$. The colored points represent the set of $\Delta\tilde{\tau}_\varphi$ and $\langle \tilde{\tau}_\varphi\rangle $ for which the Floquet exponent is $\mu_k\geq 0.01$, and therefore we have parametric resonance.  The left plot shows the stability chart for those values of the inflaton vev that correspond to heavy axions --- $m_\vartheta\in (10^{-11},10^{-5})\mpl$ --- that would make up dark matter, while the plot on the right shows higher values of $\langle \tilde{\tau}_\varphi\rangle$, which corresponds to lighter axions --- $m_\vartheta\in( 10^{-38},10^{-27})\mpl $ --- that stay relativistic until today.}
	\label{fig:stabstring}
\end{figure}

In \cref{fig:stabstring} we show the stability charts of the Hill equation in \cref{whithil} for some choice of parameters, where the red points correspond to values of ($\langle\tilde{\tau}_\varphi\rangle$, $\Delta\tilde{\tau}_\varphi$) whose solutions are exponentially unstable. We compute the instability only for $k=k_{max}$, as \cref{fig:stab} shows that this mode gives the main contribution. We require the initial displacement of the inflaton field to be lower than its vev -- that is why there is a lack of unstable solutions below the line $\Delta\tilde\tau_\varphi=\langle \tilde{\tau}_\varphi\rangle$. The two different charts correspond to two separate sets of axion masses: the left panel shows the stability chart of heavy axions ($m_\vartheta\in (10^{-11},10^{-5})\mpl$), while the one on the right shows the stability chart of light axions ($m_\vartheta\in( 10^{-38},10^{-27})\mpl $).
For both charts there are regions of exponential instability, and we expect production of both light and heavy fields.

\subsubsection{Hill equation on expanding background}

We now analyze how an expanding universe influences axion production. After inflation ends, the inflaton field behaves as a harmonic oscillator with a frequency approximately $\omega \simeq \minf$. Assuming the scale factor evolves as $a \sim t^{2/3}$, the system can be approximated using \cref{stringtoteom}.

There are some caveats to this approach. First, we must account for the transition period between the end of inflation and the onset of the oscillatory regime, during which the inflaton does not behave like a harmonic oscillator. Therefore, we set $t_0 = \frac{\pi}{2\minf}$, corresponding to the time after a quarter of one oscillation of the field $\varphi$. This provides sufficient time for the inflaton to enter the harmonic oscillatory regime.
Second, the validity of this equation of motion breaks down when the backreaction of the axions and the fragmentation of the inflaton become significant. If the inflaton loses too much energy during the preheating process and the energy density of the axions becomes comparable to that of the inflaton, the problem is better addressed with a lattice simulation.
However, backreaction and rescattering generally halt parametric resonance and the subsequent particle production. Thus, we stop our analysis once the energy density of the axions becomes comparable to that of the inflaton, i.e. when $\rho_\vartheta \simeq \mathcal{O}(10^{-1}) \rho_\varphi$. Consequently, we expect that a lattice simulation will not significantly alter the overall results.

\cref{stringtoteom} can be considered analogous to the equation of a damped harmonic oscillator with a time-dependent frequency
\begin{equation}
	\omega^2(s)=\frac{4k^2}{\minf^2}\left(\frac{1}{a}\right)^{2}-\beta\frac{\tilde{\tau}_\varphi'}{\tilde{\tau}_\varphi}\frac{s_0}{s^2}+4\frac{\Lambda^4(\tau_\varphi)}{f_\vartheta^2\minf^2}
 e^{-\langle \tilde{\tau}_\varphi\rangle}e^{-\Delta \tilde{\tau}_\varphi\cos(2s)}\,.
\end{equation}
As the inflaton oscillations dampen over time, the displacement $\Delta \tilde{\tau}\varphi(t)$ decreases as $\Delta \tilde{\tau}\varphi(t) = \Delta \tilde{\tau}\varphi(t_0) \frac{t_0}{t}$. Eventually, $\Delta \tilde{\tau}\varphi < 1$, allowing the oscillatory exponential term to be expanded, simplifying the equation to the Mathieu equation.
In this regime, parametric resonance becomes efficient when the computed Floquet exponents exceed the Hubble scale, i.e., $\mu_k > H$. However, when $\Delta \tilde{\tau}_\varphi > 1$, higher-order terms in the oscillatory exponential must be included:
\begin{equation}
	e^{-\langle \tilde{\tau}_\varphi\rangle}e^{-\Delta\tilde{\tau}_\varphi \frac{s_0}{s}\cos{(2s)}}= e^{-\langle \tilde{\tau}_\varphi\rangle}\left(1-\Delta \tilde{\tau}_\varphi \frac{s_0}{s}\cos{(2s)}+\frac{1}{2} \Delta \tilde{\tau}_\varphi^2\left(\frac{s_0}{s}\right)^2\cos^2{(2s)}+...\right)\,.
	\label{expanding}
\end{equation}
When $a=1$, in a non-expanding background, the periodic sign change of $\cos(2s)$ term temporarily offsets the suppression from $\langle \tilde{\tau}_\varphi\rangle$.
In an expanding universe ($a \neq \text{const.}$), the displacement $\Delta \tilde{\tau}_\varphi$ decreases due to cosmic expansion, making the suppression from $\langle \tilde{\tau}_\varphi\rangle$ increasingly significant over time.

To quantify resonance in an expanding background, even when $\Delta \tilde{\tau}_\varphi > 1$, we define a resonance parameter $q$:

\begin{equation}
	q=4\frac{\Lambda^4(\tau_\varphi)}{f_\vartheta^2\minf^2}\Delta \tilde{\tau}_\varphi e^{-\langle \tilde{\tau}_\varphi\rangle} 
= 4\Delta\tilde{\tau}_\varphi \frac{m_\vartheta^2}{\minf^2}
\,.
	\label{qqq}
\end{equation}
A larger $q$ corresponds to stronger parametric resonance. While $q$ scales linearly with $\Lambda^4(\tau_\varphi)$ and $\Delta \tilde{\tau}_\varphi$, it decreases quadratically with $\minf$, implying that lower inflationary scales enhance resonance. The dependence of $q$ on $\langle \tilde{\tau}_\varphi\rangle$ is non-trivial: $q$ initially increases but eventually drops sharply beyond a critical value.

\begin{figure}
	\centering
	\includegraphics[scale=0.4]{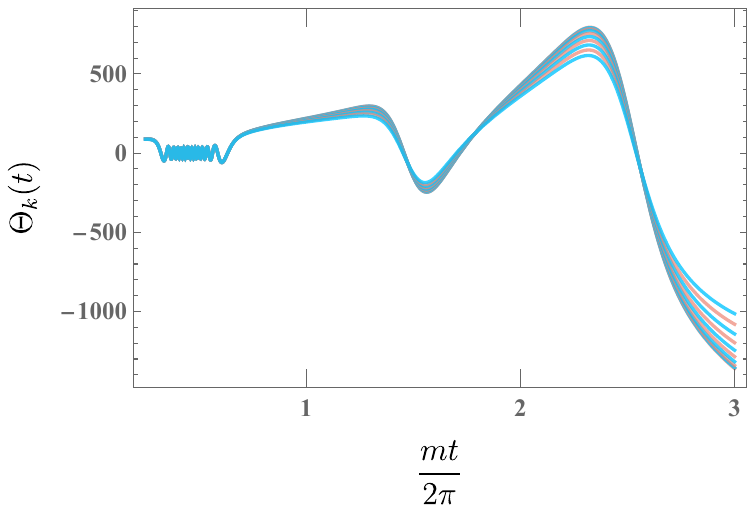}
	\includegraphics[scale=0.4]{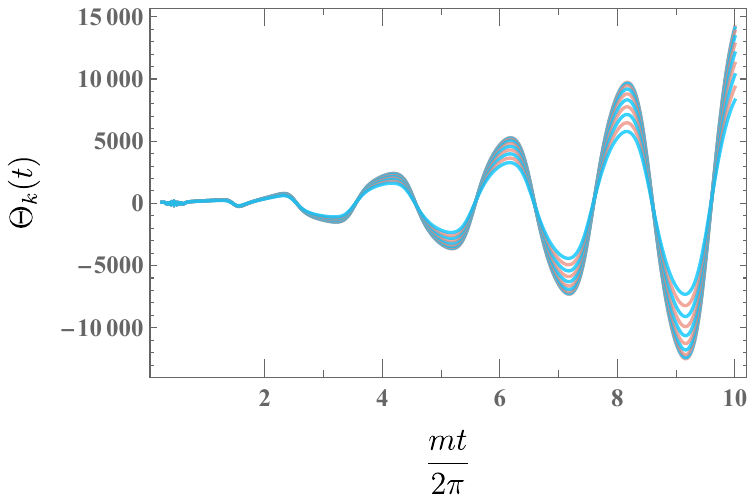}
	\includegraphics[scale=0.4]{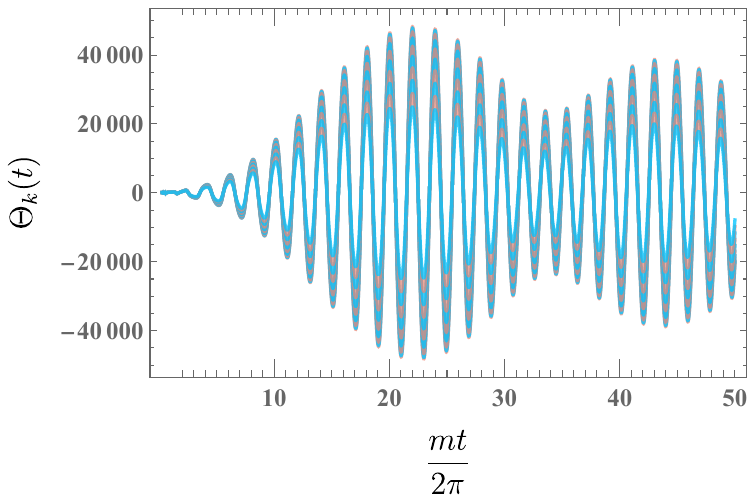}
	\includegraphics[scale=0.4]{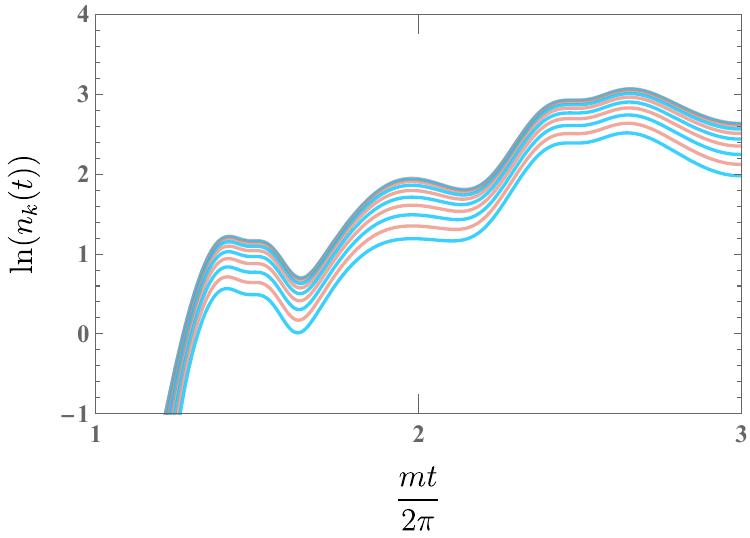}
\includegraphics[scale=0.4]{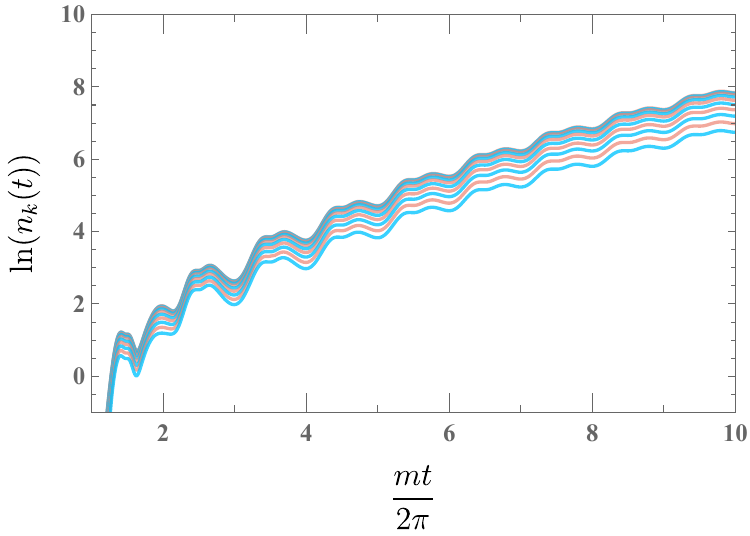}
	\includegraphics[scale=0.4]{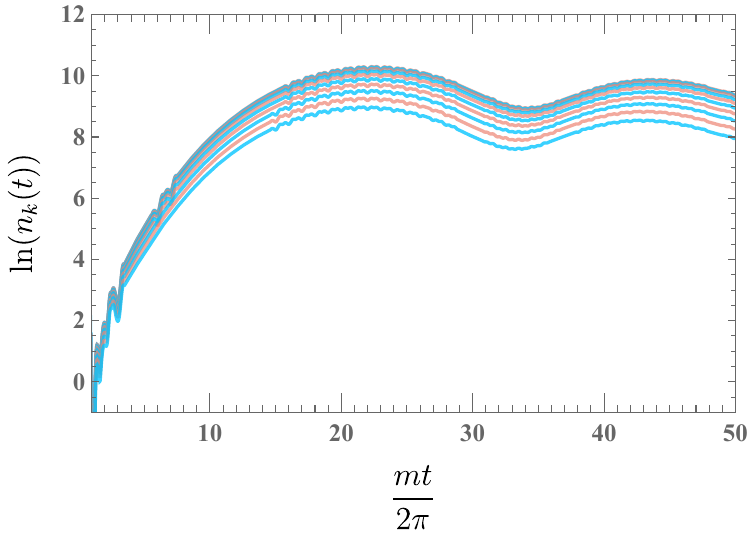}
	\caption{Different stages of parametric resonance for different modes $k$ varying around $k_{max}\simeq \minf/2$, in our theory in an expanding background with scale factor $a\sim t^{2/3}$. The values used in these plots are $\Lambda^4(\tau_\varphi) \simeq 7\times 10^{-4}$, $\minf\simeq 5\times 10^{-5}$, $\langle \tilde{\tau}_\varphi\rangle = 6.3\pi $ and $\Delta \tilde{\tau}_\varphi= 5\pi$, giving the axion mass $m_\vartheta\simeq2\times  10^{-5}\mpl$. From left to right we show longer periods of time. 
 On the top, we plot the mode evolution, while on the bottom the number of particles $n_k$. The time is shown in units of $2\pi/m$, which corresponds to the number of oscillations of the inflaton field. After around $ 20$ oscillations, the resonance ceases, and the occupation number becomes constant.} 
	\label{fig:expa}
\end{figure}

\cref{fig:expa} illustrates the growth of the modes and its number density in an expanding background. Over time, the oscillations of $\vartheta_k$ slow down, deviating from solutions derived from the Mathieu equation. Initially, axion production grows significantly, but resonance ceases after approximately 20 inflaton oscillations, as shown by the occupation number $n_k$ plateauing in the bottom panels of \cref{fig:expa}.

For the axion field to behave as radiation today, its mass must satisfy $m_\vartheta\lesssim T_{\text{CMB}}\simeq 10^{-31} \mpl $. Under this condition, the resonance parameter $ q$ is constrained as:
\begin{equation}
	q\lesssim 4 \times \frac{10^{-58}}{\minf^2} \Delta \tilde{\tau}_\varphi \left(1-\frac{1}{\langle\tilde{\tau}_\varphi\rangle}\right)\simeq 10^{-48} \Delta \tilde{\tau}_\varphi\,.
\end{equation}
In this regime, $q$ becomes extremely small, limiting the effectiveness of parametric resonance for low-mass axions. Thus, as discussed earlier, for very light axions the driving oscillations primarily arise from the kinetic mixing between the axion and the inflaton fields.

However, $q$ grows exponentially when heavier axions are considered, allowing them to serve as viable dark matter candidates.
\begin{figure}
	\centering
	\includegraphics[scale=0.59]{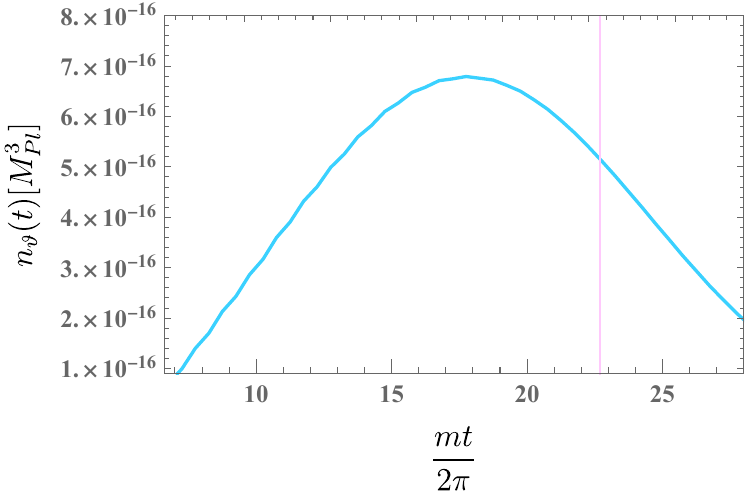}
	\includegraphics[scale=0.59]{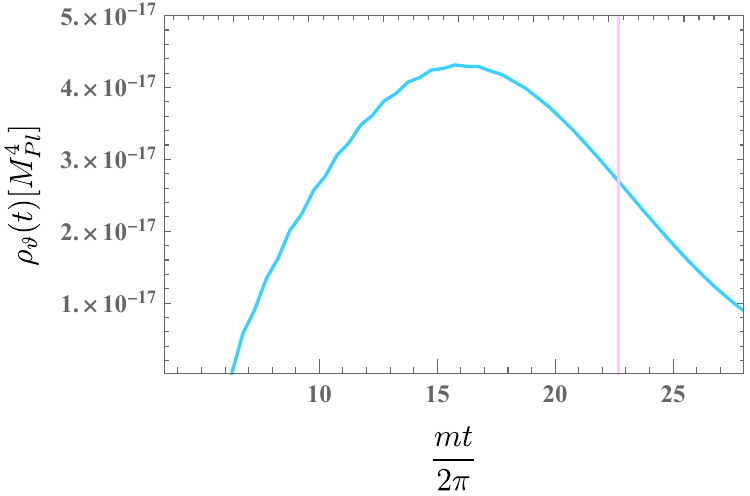}
	\caption{Number and energy density of particles created via parametric resonance for the model of \cref{fig:expa}. We see that both quantities stop growing and start redshifting before the end of parametric resonance (shown by the pink vertical line).
 }
	\label{fig:nrho}
\end{figure}
The total number and the energy density of the created particles can be computed using \cref{nchi,rhochi},  with the results shown in \cref{fig:nrho}.
During the period of parametric resonance, both quantities increase significantly, but as preheating concludes, they begin to dilute as the expansion of the universe becomes non-negligible. The particle number redshifts as $ a^{-3}$, while the energy density as $ a^{-4}$.
Interestingly, both quantities cease their growth slightly before the end of the parametric resonance, which is marked by the pink line in \cref{fig:nrho}. 
This subtle discrepancy amounts to roughly $\sim 1\%$: $n_{\vartheta}(t_{22.7})\simeq 5.2\times 10^{-16} \mpl^3$, while $n_\vartheta(t_{17.7})\simeq 6.8\times 10^{-16}\mpl^3$.

For this parameter set, the axion density parameter $\Omega_\vartheta$ can also be computed using \cref{omegachi}, revealing a significant overproduction of DM. As detailed in the previous section, such an overproduction shifts the time of matter-radiation equality to an earlier epoch. To quantify this, the time of matter-radiation equality is calculated using \cref{matdom}. Specifically, for the chosen parameters, the ratio of the scale factors is given by $\frac{a_{eq}}{a_{eq}^{\Lambda \text{CDM}}}\simeq \mathcal{O}(10^{-14})$, which is significantly less than unity.  This result implies that matter-radiation equality occurs much earlier than in the standard 
$\Lambda $CDM cosmology, leading to a universe whose evolution diverges notably from our observed one.
\begin{figure}
	\centering
	\includegraphics[scale=0.69]{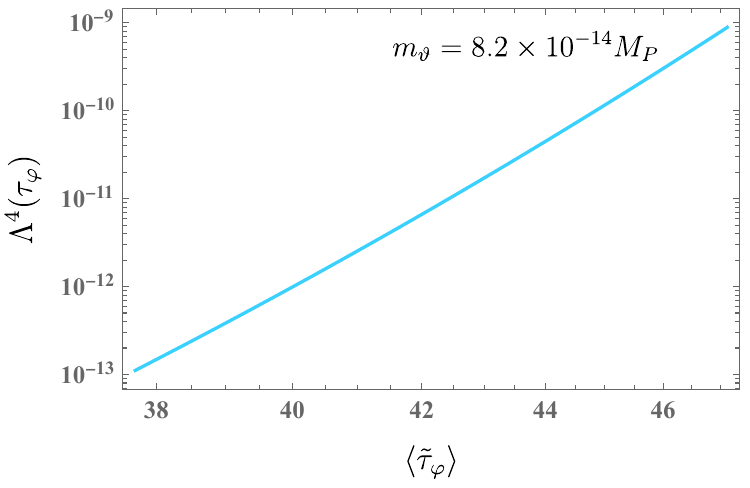}
	\caption{$\Lambda^4(\tau_\varphi)$ vs $\langle\tilde{\tau}_\varphi\rangle$ such that the mass of the axion satisfies the equality $\Omega_{\vartheta} h^2\simeq \Omega_{DM}h^2\simeq 0.12$, and makes up all of DM. } 
	\label{fig:lambdavstau}
\end{figure}
We can find the mass of the produced axions for which the axions saturate the DM bound and would make up of all DM ($\Omega_{\vartheta} h^2\simeq \Omega_{DM}h^2\simeq 0.12$). We find that this condition is satisfied for $m_\vartheta\simeq 8.2 \times 10^{-14}\mpl$. 
This result provides a constraint on the model parameters. By fixing the axion mass to this value, we can derive bounds on the remaining parameters of the model. These constraints are depicted in \cref{fig:lambdavstau}, where we plot the values of $(\tilde{\tau}_\varphi,\,\Lambda^4)$ under the requirement that axions make up the entirety of the dark matter in the universe.

\subsection{The Fate of Preheated Axions}
\label{sec:fate}
We now turn to the fate of axions produced via parametric resonance within string-inspired models. Addressing this requires examination of the possible axion decay channels into both visible and hidden sector states. This in turn depends critically on how the standard model (SM) and dark sectors are realized in the string construction. In Type IIB compactifications, for instance, the placement of the SM in extra dimensions --- specifically its relation to the four-cycles whose volume moduli drive inflation --- directly affects the coupling between the inflaton-partner axions and the SM. This, in turn, governs the axion decay rates into visible sector particles.

Without diving into the specifics of reheating channels, some general observations can be made,
Firstly, axions sourced predominantly from kinetic mixing are ultralight. Irrespective of the reheating channel, their contribution to the effective number of relativistic degrees of freedom is negligible, $\Delta N_{\rm eff} \lesssim 10^{-6}$.
For heavier axions, parametric resonance production is more efficient, but these massive axions cannot always persist through cosmic evolution without consequences. In the limit where $m_\vartheta \sim m_\varphi$, the axion becomes non-relativistic soon after production. During radiation domination, these axions redshift slower than radiation, potentially overtaking the universe's energy density and leading to overclosure.
To prevent this, the axion decay must be efficient. If the axion decays to visible sectors particles, then it simply modifies the process of reheating. If instead the axions decay to hidden sector states, it could lead to an enhancement of $\Delta N_{\rm eff}$. We now enumerate the most obvious decay channels for the axions produced during preheating. 

Axions couple to gauge fields via the usual Chern-Simons coupling 
\begin{equation}
    \mathcal{L}\supset -\frac{g_{\vartheta    \gamma\gamma}}{4}\vartheta F\widetilde{F}\,\fstop
\end{equation}
where $g_{\vartheta   \gamma\gamma }\sim \frac{1}{f_\vartheta}$.
Therefore, the axion can decay into two gauge bosons with a decay rate 
\begin{equation}
    \Gamma_{\vartheta\to \gamma\gamma }=\frac{g_{\vartheta   \gamma\gamma}^2 m_\vartheta^3}{64\pi}\,.
\end{equation}
If the PR axions couple directly to the visible sector, the above allows a direct decay into visible sector gauge bosons. Alternatively, the gauge boson could be part of the hidden sector. In the context of type IIB, this can occur when a stack of D7-branes is supported on a blow-up cycle. In such a situation, the PR axions would eventually turn into a component of dark radiation. However, this is a difficult scenario to make consistent with reheating. Via $\mathcal{N}=1$ supersymmetry, one expects that the scalar partner of the PR axion couples to the dark gauge bosons via an operator $\varphi F^2$. If this scalar field is the inflaton, then this operator yields a direct decay channel in dark radiation, disrupting the latter process of reheating. For fibred CY compactifications in fiber inflation models, this issue was first raised in~\cite{Anguelova:2009ht}. Excessive dark radiation from these decays places significant constraints on such constructions. Current studies often avoid including this  $U(1)$ due to these challenges. More generally, it is difficult to construct scenarios where the axion interacts with a hidden sector while the corresponding saxion remains decoupled. In most cases, the inflaton decays perturbatively into the dark sector, saturating or exceeding the bounds on dark radiation.

\begin{figure}
	 \centering
    \hspace{0.8cm}
    \includegraphics[scale=1.2]{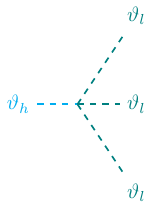}\hfill
    \raisebox{9mm}{\includegraphics[scale=1.2]{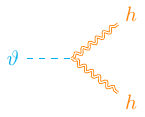}}\hfill
      \raisebox{2mm}{\includegraphics[scale=1.2]{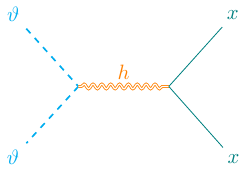}}
 \hspace{0.8cm}
	\caption{Feynman diagrams of the relevant processes for Case II-c where the produced axion is heavy and not stable. On the left, we depict the three body decay of the heavy axion into the light ones (green dashed lines). The middle diagram shows the  heavy axion (blue dashed lines) decaying into gravitons (double squiggly orange lines). The right diagram instead shows the gravity-mediated scattering of the heavy axion into every other d.o.f. of the theory, which we label as $x$ (continued green line). } 
	\label{fig:feynman_diagr}
\end{figure}

A similar decay channel occurs via the gravitational Chern-Simons coupling of the axion. The relevant operator is
 \begin{equation}
        \mathcal{L}\supset \frac{g_{\vartheta    hh}}{4}\vartheta R\widetilde{R}\,,
    \end{equation}
with $g_{\vartheta    hh}\sim 1/f_\vartheta$ which arises naturally in string theories~\cite{Witten:1984dg,Choi:1985je,Choi:1985bz,Antoniadis:1993jc,Banks:1996ss,Choi:1997an,Choi:1999zy}, see \cref{app:gravitationalCS}. The Feynman diagram of this process is shown as the middle image of \cref{fig:feynman_diagr}. This interaction is characterized by the decay rate
~\cite{Delbourgo:2000nq,Alonzo-Artiles:2021mym,Ema:2021fdz,Yang:2023vwm}
 \begin{equation}
        \Gamma_{\vartheta  \rightarrow hh} = \frac{g_{\vartheta    hh}^2m_\vartheta^7}{512\pi \mpl^4}\,.
 \end{equation}
We can give an estimate of the order of magnitude by using the coupling as in \cref{ahhcoupling}:
\begin{equation}
     \Gamma_{\vartheta  \rightarrow hh}=\left(\frac{N}{384 \pi^2 f_\vartheta}\right)^2\frac{m_\vartheta^7}{512\pi \mpl^4} \sim \mathcal{O}(10^{-11})\frac{m_\vartheta^7}{f_\vartheta^2\mpl^4}\sim \mathcal{O}(10^{-7})\frac{m_\vartheta^7}{\mpl^6}\,,
\end{equation}
where in the last equality we used $f_\vartheta\sim 10^{-2}\mpl$. While gravitons produced this way would constitute a contribution to $\Delta N_{\eff}$, the decay rate is so small as to be negligible. If the PR axions have no decay channel other than to gravitons, we can approximate them as stable for the purposes of their cosmological impact.

Most inflationary models in string compactifications are characterized by the presence of more than one axion.
Via the kinetic mixing of the heavy and light axions (respectively $\vartheta_h$ and $\vartheta_l$), one obtains the following coupling
    \begin{equation}
        \mathcal{L}\supset \lambda \vartheta_h \vartheta_l^3 \,.
    \end{equation}
This represents the possibility of a three body decay (see the left Feynman diagram in \cref{fig:feynman_diagr}) that,
in the limit $m_{\vartheta_l}\rightarrow 0 $, has a decay rate
    \begin{equation}
        \Gamma_{\vartheta_h\rightarrow \vartheta_l\vartheta_l\vartheta_l} = \frac{9\lambda^2m_{\vartheta_h}}{2(4\pi)^3}\,.
    \end{equation}
However, the coupling of the kinetic mixing between the axions depends on the mass of the light axion is $\lambda\sim \frac{m_{\vartheta_l}^2}{f_\vartheta^2}$. Indeed, when the mass of the light axion is very suppressed, the decay rate will be as well.

The produced axions have a cross-section to produce all other particles of the theory via intermediary gravitons. We call these particles $x$, and they represent both the visible and the hidden sectors. The Feynman diagram of this interaction is shown as the right figure of \cref{fig:feynman_diagr}. 
Such an idea was also used for \textit{gravitational reheating} in~\cite{Mambrini:2021zpp,Barman:2021ugy,Haque:2022kez}. 
The scattering rate, in the case where $m_x\to0$ reads 
\begin{equation}
    \Gamma_{\vartheta  \vartheta\to xx}\sim \frac{1}{1024 \pi}\frac{ n_\vartheta m_\vartheta^2}{\mpl^4}=\frac{1}{1024 \pi}\frac{ \rho_\vartheta m_\vartheta}{\mpl^4}\,.
\end{equation}
To estimate this scattering rate, we can relate the axion properties to those of the inflaton field: in our analysis we stop the resonance when the energy density of the produced axions is at most $\rho_\vartheta\sim 0.1\,  \rho_\varphi=0.1\,\times  3H^2 \mpl^2$, which at the end of inflation reads $\rho_\varphi=3\minf^2\mpl^2$. Furthermore, we are producing axions whose masses are not many orders of magnitudes away from that of the inflaton field. 
Having postulated this, the scattering rate can be roughly estimated as 
\begin{equation}
    \Gamma_{\vartheta  \vartheta\to xx}\sim\mathcal{O}(10^{-5})\,\frac{m_\vartheta^3}{\mpl^2}\,.
\end{equation}
We can therefore write the following set of equalities, relating the decay rates to that of the axion photon decay $\Gamma_{\vartheta\to \gamma\gamma}$:
\begin{equation}
   \Gamma_{\vartheta\to hh}\simeq \mathcal{O}(10^{-2})\,\Gamma_{\vartheta\vartheta\to xx}\frac{m_\vartheta^4}{\mpl^4}\sim \mathcal{O}(10^{-5})\, \Gamma_{\vartheta\to \gamma\gamma} \frac{m_\vartheta^4}{\mpl^4}\frac{f_\vartheta^2}{\mpl^2} \,.
\end{equation}
Taking $f_\vartheta\sim 10^{-2}\mpl$, and specifying the mass of the axion to be lower than $H$, then by taking $H\sim 10^{-5}\mpl$, the biggest values we obtain result in $\Gamma_{\vartheta\to \gamma\gamma}\sim 10^{-13}\mpl$ and
\begin{equation}
   \Gamma_{\vartheta  \to hh}\sim \mathcal{O}(10^{-22})\,\Gamma_{\vartheta  \vartheta\to xx}\sim \mathcal{O}(10^{-29})\, \Gamma_{\vartheta  \to \gamma\gamma} \sim 10^{-42}\mpl\,.
\end{equation}
It is clear from this that the main decay channel, if it is present, will be the axion photon decay, as the gravitons-mediated interactions are very suppressed. 
However, if there is no axion-photon coupling, and the only possible interactions are the gravitational ones, the scattering of axions into $x$ particles will be the predominant channel.
Indeed, the gravitational axion scattering rate is greater than the axion-graviton decay rate. 
This happens because the scattering rate is enhanced by the number density of the axion particles, which grows during preheating, and overcomes the suppression factor ($\frac{1}{\mpl^4}$)  common between the two rates.

We note that this hierarchy is valid at early times, when we can neglect the dilution of the energy densities due to expansion. At later times, the redshifting of the axion energy density will reverse the hierarchy.

\section{Application: Fibre Inflation}\label{sec:FibreInf}

We now specialize to a particular realization of string-inspired inflation models in the context of fibre inflation models embedded in the Large Volume Scenario (LVS) compactifications.

\subsection{Review of LVS \& Fibre Inflation}

We will consider a  fibred CY 3-fold with volume 
\begin{equation}
    \mc{V} = \alpha \left(\sqrt{\tau_1}\tau_2-\gamma\tau_3^{3/2}\right)\,.
\end{equation}
Crucially, the K\"ahler moduli and their axionic partners are stabilized by quantum corrections to $K$ and $W$. The K\"{a}hler potential including perturbative corrections is given by 
\begin{equation}
    \begin{aligned}
        K &=\kah(g_s,\langle z_a \rangle)+ K_{0} +  K_{\alpha^\prime} + K_{g_s}\coma\\
        K_0 &= -2\ln(\mc{V})\,,\\
        K_{\alpha^\prime} &=  - \hat{\xi}/\mc{V}\,.
    \end{aligned}    
\end{equation}
where $K_{0}$ is the tree-level K\"{a}hler potential that depends on the volume of $X_6$, $\mc{V}$, while 
$\kah(g_s,\langle z_a \rangle)$ encloses the contributions from $g_s$ and the stabilized complex structure moduli $z_a$, and contributes an overall factor to the potential. The corrections come from higher derivatives ($K_{\alpha^\prime}$) and string loops ($ K_{g_s}$). The $\alpha^\prime$ correction depends on the Euler number of $X_6$, $\chi(X_6)$, via the parameter 
$\hat{\xi} := - \frac{\zeta(3)\chi(X_6)}{2(2\pi)^3g_s^{3/2}}$ \cite{Becker:2002nn} while loop corrections depend on the vacuum value of the complex structure moduli which are stabilized at tree-level by fluxes \cite{vonGersdorff:2005bf,Berg:2005ja,Cicoli:2007xp}. Additionally, one can also consider higher superspace-derivative corrections, which might be relevant for fibre inflation~\cite{Ciupke:2015msa,Broy:2015zba,Kallosh:2017wku}.\footnote{We omit a detailed treatment of both higher derivative and string loop corrections as their precise form will not affect the analysis in this work.} 

The non-perturbative contributions to $W$ that we consider arise from D7-branes or ED3s wrapping 4-dimensional cycles in the CY. Those take the form
\begin{equation}
    W=W_0+\sum_{i}A_i e^{-a_i  T_i}\,,
\end{equation}
where $A_i$ are the one-loop Pfaffian (that can depend on the complex structure moduli, the axiodilaton, and the brane moduli, which we take to be stabilized such that $A_i$ is effectively a real constant), and $a=2\pi $ for ED3s or $a=2\pi/N $ for D7-branes, where $N$ is the dual Coxeter number of the gauge group generated by the stack of branes.

For the purpose of computing an inflationary potential in the next sections, we focus on the large volume limit of the effective theory: in this way we can recast the potential in inverse powers of $\mc{V}$. At first order, which corresponds to $\mc{O}(\mc{V}^{-3})$ , we have the LVS potential \cite{Balasubramanian:2005zx}
\begin{equation}
\label{VLVS}
    V_{\text{LVS}}\cdot e^{-\kah}=\frac{8 a_3^2 A_3^2}{3\alpha\gamma} \frac{\sqrt{\tau_3}\, e^{-2 a_3 \tau_3}}{\mathcal{V}}+4 a_3 A_3 W_0 \frac{\tau_3 e^{-a_3 \tau_3}}{\mathcal{V}^2}\cos(a_3\theta_3)+\frac{3 \hat{\xi } W_0^2}{4 \mathcal{V}^3}+\delta V_{\rm uplift}\,.
\end{equation}
At this order, the axion $\theta_3$, small cycle modulus and the overall volume are stabilized respectively at
\begin{equation}
    \theta_3=\frac{\pi}{a_3}\coma \tau_3=\left(\frac{2\alpha  \gamma }{\hat{\xi}}\right)^{-2/3}\coma \mc{V}=\frac{3 \alpha  \gamma  \sqrt{\tau_3}\,W_0 }{4 a_3A_3}\,e^{a_3 \tau_3}\,,
\end{equation}
and the vacuum is a SUSY-breaking AdS minimum, which can be tuned to near-zero CC of either sign by the uplift term $\delta V_{\rm uplift}$ (for a discussion of the possible sources of uplifting see e.g.~\cite{McAllister:2023vgy}). The fibre volume modulus $\tau_1$ obtains a vev due to the K\"{a}hler potential string loop corrections $K_{g_s}$. The relevant terms in the potential arise at $\mc{O}(\mc{V}^{-10/3})$ and are
\begin{equation}
    V \supset \bigg(g_s^2 \frac{A}{\tau_1^2} - \frac{B}{\vol\sqrt{\tau_1}} + g_s^2\frac{C\tau_1}{\vol^2}\bigg) \frac{W_0^2}{\vol^2}\,,
\label{eq:inflpot}
\end{equation}
where we used $\tau_2\simeq \langle\mc{V}\rangle/\alpha\sqrt{\tau_1}$. We can recast~\cref{eq:inflpot} into the inflaton potential by defining the canonically normalized field~\cite{Cicoli:2008gp}
\begin{equation}\label{eq:FibreInfCanNorm}
    \varphi\equiv \frac{\sqrt{3}}{2}\ln \tau_1\,,
\end{equation}
and considering its shift from the vacuum value, $\varphi=\langle\varphi\rangle+\hat{\varphi}$, such that $V_\infl(\langle\varphi\rangle)=0$,
the inflaton potential from fibre inflation reads
\begin{equation}
 \begin{aligned}
     V_\infl&=V_{\text{LVS}}+\frac{W_0^2}{\mathcal{V}^2}\left(g_s^2A e^{-2\kappa \hat\varphi}-\frac{B}{\mathcal{V}}e^{-\kappa\hat\varphi/2}+\frac{g_s^2C}{\mathcal{V}^2}e^{\kappa\hat\varphi}\right)\,.
\end{aligned}
\label{eq:canoninfpot}
\end{equation}
Here, we defined $A=A_{\rm loop} e^{-2\kappa\langle\varphi\rangle}$, $B=B_{\rm loop} e^{-\kappa\langle\varphi\rangle/2}$ and $C=C_{\rm loop} e^{\kappa\langle\varphi\rangle}$ in terms of $A_{\rm loop}$, $B_{\rm loop}$, and $C_{\rm loop}$ which contain the string 1-loop corrections to the K\"{a}hler potential of the K\"{a}hler moduli.

The $\theta_1$ and $\theta_2$ axions obtain their masses only at $\mc{O}(\mc{V}^{-\frac43}e^{-\mc{V}^{\frac23}})$ from the terms
\begin{align}
    V&\supset e^{K_0}\bigg(K_0^{\bar{T}_1T_1}\bar{F}_1 F_1 +K_0^{\bar{T}_2T_2}\bar{F}_2 F_2 \bigg)\\
      & \simeq \frac{4a_1A_1W_0}{\vol_0^2} \tau_1 e^{-a_1\tau_1}\cos(a_1\theta_1) + \frac{4a_2A_2W_0}{\vol_0^2} \tau_2e^{-a_2\tau_2}\cos(a_2\theta_2)   \,.
\end{align}
Thus the axions have vanishing vevs $\langle \theta_1\rangle=\langle \theta_2\rangle = 0$. Akin to the inflaton, we will need the canonically normalized axions to study parametric resonance. Following the procedure discussion in~\cref{sec:string},
the K\"ahler metric $g_{ij}=2\frac{\partial^2 K}{\partial T^i \partial \bar{T}^j}$ at leading order reads
\begin{equation}
g_{ij}\simeq 2\left(
\begin{array}{ccc}
 \frac{1}{4 \tau_1^2} & 0 & 0 \\
 0 & \frac{\alpha ^2 \tau_1}{2 \mathcal{V}^2} & 0 \\
 0 & 0 & \frac{3 \alpha  \gamma}{8 \sqrt{\tau_3} \mathcal{V}} \\
\end{array}
\right)\,,
\end{equation}
where we replaced the modulus $\tau_2$ with its vev. Thus we find decay constants
\begin{equation}
    f_1=\frac{1}{\sqrt{2}\,a_1\tau_1}\quad\mbox{and}\quad f_2=\frac{\alpha}{a_2\tau_2}\,, 
\end{equation}
and $a_i\theta_i=\vartheta_i/f_i$. For the axionic partner of the inflaton, the canonically normalized potential is
\begin{equation}
 -V_{\text{ax}}\simeq 
 \frac{8 a_1  |A_1 W_0|\tau_1}{\mathcal{V}^2}e^{-a_1\tau_1}\cos{ (\vartheta_1/f_1)}\,.
    \label{firts_pot}
\end{equation}
%


\subsection{The Visible Sector and Perturbative Reheating in Fibre Inflation}

The above ingredients stabilize the compactification and realize inflationary physics. From here, we can study the preheating of axions in the model. Before doing so, we first review certain requirements for viable fibre inflation models. These requirements illuminate the visible and hidden sector content of the EFT, which is critical for predicting the eventual fate of preheated axions. This step is also essential for constructing a complete cosmological model, as the compactification must feature a sector that mimics the minimal supersymmetric standard model (MSSM) or an extension thereof, as well as a mechanism to reheat this MSSM-like sector.

The interplay between the SM-like sector, fibre inflation, and axion physics can lead to two main possible outcomes for the inflaton-partner axion:

\begin{enumerate}[I-]
    \item If the SM sector resides on one of the fibration four-cycles that drive inflation, this setup enforces the absence of stringy instantons on that cycle, leaving the partnered axion extremely light. Due to the shared four-cycle, this axion may interact with the SM gauge fields. In such a scenario, the axion could manifest as an ultra-light Cosmic Axion Background (CaB), contributing negligibly to dark radiation with a maximal $\Delta N_{\rm eff}\lesssim10^{-6}$. If the axion decays into SM particles, such as photons, no observable axion relic would remain.

    \item If the SM is instead located on an additional, smaller blow-up four-cycle, the inflaton-partner axion becomes sequestered from the SM and forms part of the dark sector. This case leads to several sub-scenarios:
 \begin{enumerate}[a.]
\item  A super-sequestered axion scenario, where the axion decays solely within the dark sector. If the axion is sufficiently light, it behaves as a CaB; otherwise, it risks overproduction, contributing excessively to dark matter.
\item  The axion decays into heavier dark sector states, such as those associated with a condensing gauge group. The outcome mirrors the super-sequestered case (II-a).
\item Couplings between axions and additional light sectors allow the axion to decay further. In the absence of such couplings, the only remaining interaction would be gravitational.
    \end{enumerate}
\end{enumerate}
To better understand these scenarios, we now outline the necessary conditions for a consistent fibre inflation framework and reheating mechanisms in these setups.
The MSSM-like sector can arise from either D7-branes wrapping 4-cycles of $\widetilde{X}_6$ or D3-branes localized at singularities. The coexistence of a SM-like particle physics sector and successful fibre inflation within the same CY orientifold compactification places certain constraints on the total setup: 
\begin{itemize}
\item Fibre inflation requires a certain set of string loop corrections to generate the inflationary scalar potential. 

One generic way often studied in literature to ensure this condition consists of putting D7-brane stacks on the $\tau_1$ and $\tau_2$ fibration 4-cycles~\cite{Cicoli:2010ha}.  This condition becomes significantly less constraining if some of the loop corrections arise from 10D bulk loops of closed strings as argued in~\cite{Gao:2022uop}.
\item The LVS mechanism of K\"{a}hler moduli stabilization underlying a viable setup for fibre inflation requires the presence of a non-perturbative effect on the LVS blow-up 4-cycle responsible for the stabilization of the CY volume. 

This condition can be satisfied by either wrapping a Euclidean D3-brane (producing an ED3 instanton) or a small D7-brane stack (producing gaugino condensation) on the LVS blowup 4-cycle. Using the ED3 variant is unfeasible as this results in LVS stabilization of the overall CY volume at a value which renders the scale of the fibre inflation scalar potential incompatible with CMB normalization~\cite{Cicoli:2010ha}.
 Hence, the first possibility to implement this and the above condition unavoidably give rise to hidden sectors. 

 \item A SM-like particle physics sector only works if the 4-cycle carrying the SM sector 7-branes carries no brane instantons and does not intersect with any other instanton-generating 4-cycle in the CY orientifold.

 This condition in turn requires~\cite{Blumenhagen:2007sm,Cicoli:2017axo} a judicious choice of D7-brane gauge fluxes to avoid the unwanted intersections and/or to de-rigidify the fibre or base 4-cycle in case the SM sector 7-branes were wrapped on one of them. Finally, there is the phenomenological requirement~\cite{Cicoli:2010ha} of putting the SM sector, when realized on 7-branes, on a small 4-cycle (typically this limits placement to one of the blow-ups) to avoid too weak SM sector gauge couplings.
\end{itemize}
 
Let us now briefly review the literature on reheating in fibre inflation models~\cite{Cicoli:2008gp,Cicoli:2010yj,Cicoli:2012aq,Higaki:2012ar,Angus:2014bia,Allahverdi:2014ppa,Cicoli:2018cgu,Cicoli:2022uqa}. Reheating mechanisms vary depending on whether the SM sector arises from D7-branes or D3-branes. If the MSSM-like sector arises from D7-branes on the fibre divisor, the inflaton directly couples to the visible sector. This leads to dominant inflaton decay modes into visible gauge bosons and Higgs degrees of freedom. On the other hand, for a visible sector realized by D3-branes at a singularity, reheating is more challenging due to the absence of direct couplings. Modifications to the Giudice-Masiero term in the Kähler potential have been proposed~\cite{Cicoli:2022uqa} to facilitate reheating while avoiding excessive dark radiation production.
These terms read 
\begin{equation}
    \begin{aligned}
    \label{GiudiceMasiero}
\frac{K}{\mpl^2} \supset & \frac{H_u \bar{H}_u}{\left(T_1+\bar{T}_1\right)^{y_1}\left(T_2+\bar{T}_2\right)^{y_2}}+\frac{H_d \bar{H}_d}{\left(T_1+\bar{T}_1\right)^{w_1}\left(T_2+\bar{T}_2\right)^{w_2}} +\\
& +\frac{Z H_u H_d+\text { h.c. }}{\left(T_1+\bar{T}_1\right)^{k_1}\left(T_2+\bar{T}_2\right)^{k_2}}\,,
\end{aligned}
\end{equation}
where $y_1+y_2=w_1+w_2=1$ and $k_i=y_i+w_i/2$ for $i=1,2$, $H_u$ and $H_d$ are the MSSM Higgs doublets and $Z$ represents a bilinear coupling of the Higgs fields. 
The Giudice-Masiero mechanism traditionally introduces terms that couple the moduli to visible-sector fields, enabling the generation of soft supersymmetry-breaking terms in supergravity. However, in this context, it serves an additional purpose: facilitating reheating by allowing the inflaton to decay into MSSM fields indirectly. 

The Giudice-Masiero terms also affect the dynamics of axions. In cases where the SM sector is on D3-branes, the axion, being the imaginary part of the fibre Kähler modulus, does not directly couple to visible-sector fields. The introduction of the Giudice-Masiero terms provides an indirect decay channel for the inflaton as well as for its partner axion, ensuring that the energy initially stored in the axion field can be transferred to MSSM particles, such as Higgs bosons or gauge fields. As we will see, this interaction is crucial if one wants to mitigate overclosure: in the absence of such terms, the axion could dominate the energy budget of the Universe, either as a component of dark radiation or dark matter.

\subsection{Preheating in Fibre Inflation}
We now turn to preheating in fibre inflation models. We will take the inflationary potential given in~\cref{eq:canoninfpot} and consider the non-perturbative production of $\vartheta \equiv \vartheta_1$ axions, whose potential is given in~\cref{firts_pot}. After Fourier transforming the $\vartheta$ axion field, the relevant system of equations to be solved is as in \cref{floquet}
\begin{equation}
    \begin{dcases}
 \ddot{\hat\varphi}+3H\dot{
\hat\varphi}+\left(\frac{\partial V}{\partial \hat\varphi}\right)  =0\,, \\
 \ddot{\vartheta}_{k}+ \left(3H-2\frac{\dot{\tau_1}}{\tau_1}\right)\dot{\vartheta}_k+\left(k^2+\frac{\partial^2 V}{\partial \vartheta^2}\right)  \vartheta_{k}=0\,,
\end{dcases}
\end{equation}
which can be rewritten in terms of $\Theta_k=a^{3/2} \vartheta_k$, with $'=\frac{d}{ds}$ as 
\begin{equation}
    \begin{dcases}
 \hat\varphi''+3 \frac{2}{m_\varphi}H
\hat\varphi'+\frac{4}{m_\varphi^2}\left(\frac{\partial V}{\partial \hat\varphi}\right) =0\,, \\
 \Theta_{k}''+ 2\hat\varphi' \Theta_k'+ \left(\frac{4}{\minf^2}k^2-\frac{\hat\varphi'}{a^{3/2}}+64  \frac{|A_1 W_0|}{\minf^2\mathcal{V}^2}\langle\tilde{\tau}_1\rangle^3 e^{\frac{2}{\sqrt{3}}\hat\varphi}e^{-\langle\tilde{\tau}_1\rangle e^{\frac{2}{\sqrt{3}}\hat\varphi}}\right)\Theta_k =0\,,
\end{dcases}
\end{equation}
Here, we again define as in \eqref{eq:FibreInfCanNorm} the canonically normalized inflaton as $\varphi\equiv \frac{\sqrt{3}}{2}\ln \tau_1 $ and split this as $\varphi=\langle\varphi\rangle+\hat\varphi$ into the vacuum expectation value $\langle\varphi\rangle=\frac{\sqrt{3}}{2}\ln \langle\tau_1\rangle$ at the stabilized minimum and the displacement $\hat\varphi=\frac{\sqrt{3}}{2}\ln \frac{\tau_1}{\langle\tau_1\rangle}$.

\begin{figure}[t!]
	\centering
	\includegraphics[scale=0.4]{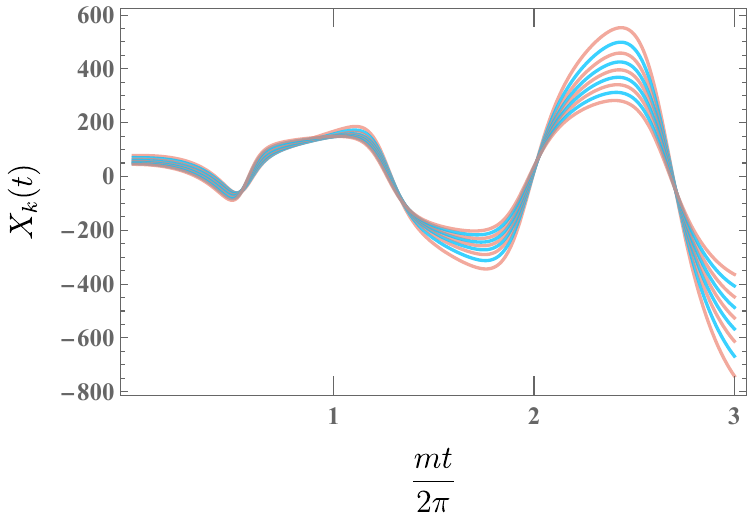}
	\includegraphics[scale=0.4]{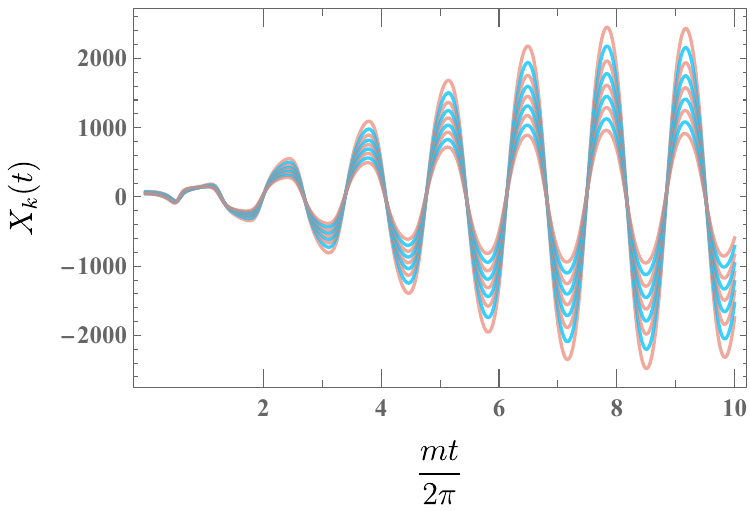}
	\includegraphics[scale=0.4]{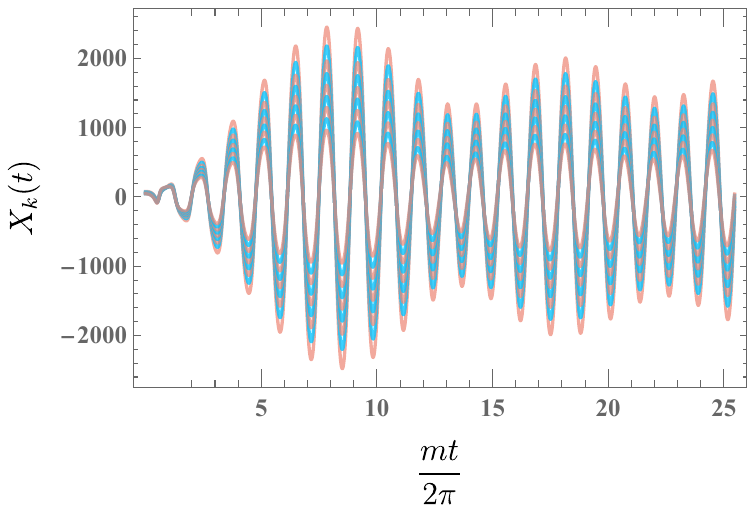}
	\includegraphics[scale=0.4]{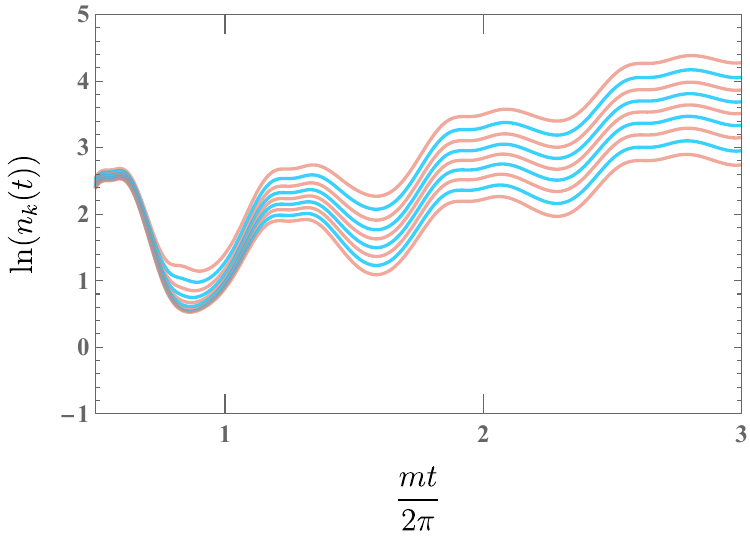}
	\includegraphics[scale=0.4]{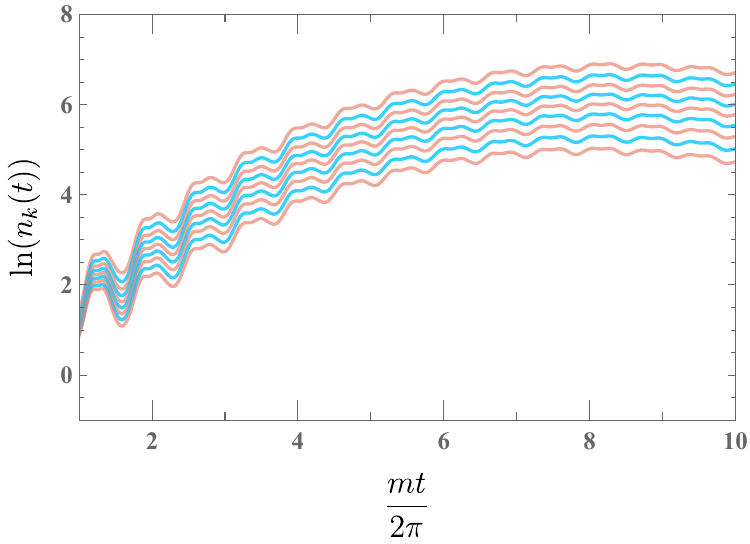}
	\includegraphics[scale=0.4]{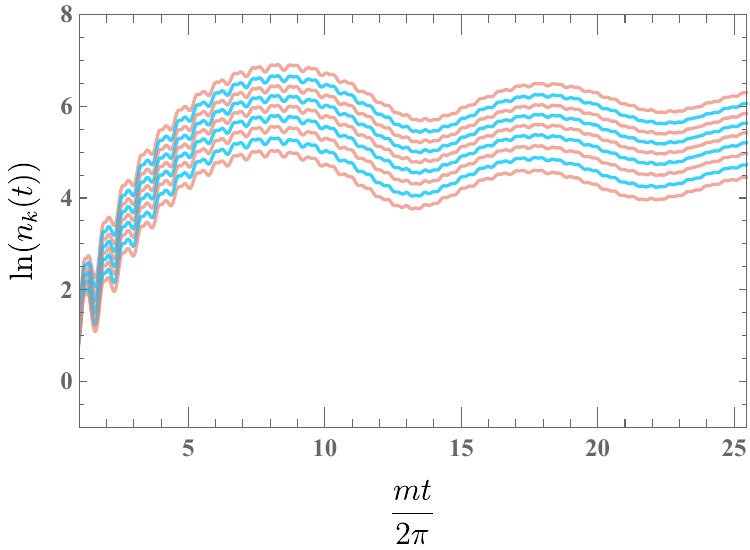}
	\caption{Different stages of parametric resonance in our theory in an expanding universe, for the values in \cref{table}, inflaton mass $\minf\simeq 7.7 10^{-4}\mpl$, and axion mass $m_\vartheta\simeq 6.77\times 10^{-5}\mpl $. From left to right, we show longer periods of time. On the top, we display the mode evolution, while on the bottom the number of particles $n_k$. The time is shown in units of $2\pi/m$, which corresponds to the number of oscillations of the inflaton field. After around $10$ oscillations, resonance ceases and the occupation number becomes constant.} 
	\label{fig:expaF}
\end{figure}

We plot in \cref{fig:expaF} the mode functions and the evolution of the particle number density. 
We can again compute the reheating temperature, and the DM abundance. 
We find also here that $\Omega_{\vartheta}>\Omega_{DM}$.
During the evolution of the universe, the produced axion particles overcome the radiation produced during reheating before $a_{eq}^{\Lambda \text{CDM}}$, and therefore would change the history of the universe.

We note that one should also consider the possible self-resonance from the inflaton field induced on itself. 
The equations one needs to solve will look fairly similar:
\begin{equation}
    \begin{dcases}
 \hat\varphi''+3 \frac{2}{m_\varphi}H
\hat\varphi'+\frac{4}{m_\varphi^2}\left(\frac{\partial V}{\partial \hat\varphi}\right) =0\,, \\
\hat\varphi_{k}''+ 2\hat\varphi' \hat\varphi_k'+ \left[\frac{4}{\minf^2}k^2-\frac{\hat\varphi'}{a^{3/2}}+\frac{4}{\minf^2} \frac{\partial^2V_\infl}{\partial\hat\varphi^2}\right. \\
\qquad\qquad\qquad \left.+64 \frac{|A_1 W_0|}{\minf^2\mathcal{V}^2}\langle\tilde{\tau}_1\rangle^2  \left(\langle\tilde{\tau}_1\rangle e^{\frac{2}{\sqrt{3}}\hat\varphi}-2\right)e^{-\langle\tilde{\tau}_1\rangle e^{\frac{2}{\sqrt{3}}\hat\varphi}}\right]\hat\varphi_k =0\,, 
\end{dcases}
\end{equation}
We check that this produces a negligible amount of parametric self-resonance.
Let us consider the following parameters
\begin{equation}
\label{table}
\begin{array}{|c|c|c|c|c|c|c|c|c|c|} 
\hline
\xi  &a_3&\,A_3\, &\,W_0\, &g_s &\alpha&\gamma& \left\langle\tau_3\right\rangle &\left\langle\tau_1\right\rangle&\langle\mathcal{V}\rangle \\
\hline
0.894 & \pi / 4& \,1\, & 10\,& 0.28 & 0.25&3.01&3.09&6.28&936\\
\hline
\end{array}
\end{equation}
By taking the second derivative of \cref{firts_pot} we see that the axion mass is 
\begin{equation}
    m_{\vartheta_i}\simeq  \frac{8 a_i |A_i W_0|}{\mathcal{V}^2 f_i^2}\tau_i e^{-a_i \tau_i}\,.
\end{equation}
For concreteness, we take the approximated value for the vev of the fibre modulus $\tau_1$, stabilized as $\langle \tau_1\rangle\simeq g_s^{4/3} \left(\frac{4A}{B}\langle\mathcal{V}\rangle\right)^{2/3} \sim 6.28$, then the base modulus $\tau_2$ acquires a vev $\langle\tau_2\rangle=\frac{\langle\mathcal{V}\rangle}{\alpha\sqrt{\langle\tau_1\rangle}}\sim \mathcal{O}(10^3)$. The mass of the base axion will be exponentially suppressed by $\tau_2$, and therefore we can safely approximate $m_{\vartheta_2}\to0$.  
We note that, if one of the axions is heavy, the other one has to be very light. This is because the volume modulus is fixed, and the fibre and base moduli vary accordingly.

In order to keep track of the complete evolution of the inflaton field from the end of inflation on, we keep the full potential.
We note that the validity of this analysis is limited once backreaction of the axions and fragmentation of the inflaton become important. 
If the inflaton loses too much energy in the preheating process, and the energy density of the axions becomes comparable to that of the inflaton, a lattice simulation would be in place. 
However, the generic effects of backreaction and rescattering are to stop parametric resonance, and subsequently to stop particle production. Therefore, we stop our analysis once the energy density of the axions becomes comparable to that of the inflaton, i.e.  $\rho_\vartheta\simeq \rho_\varphi$, and we expect that a lattice simulation will not change much the overall result. While this regime of strong backreaction is beyond the scope of our work, we note that a full treatment with lattice simulations may show further non-perturbative phenomena such as the formation of oscillons~\cite{Fodor:2019ftc} or axion stars~\cite{Zhang:2018slz} (i.e. a Bose-Einstein condensate).


\subsection{Constraints on Fibre Inflation}

%
\begin{figure}[t!]
	\centering
 \includegraphics[scale=0.9]{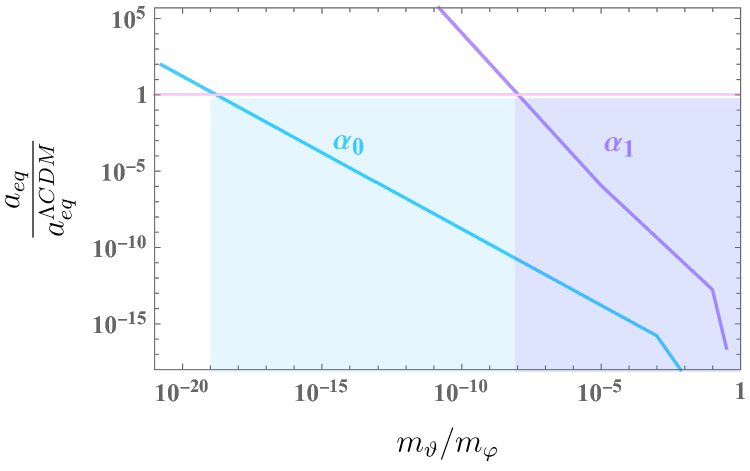}
	\caption{Blue: Case II-a, idealized case where the axion does not decay. Purple: Case II-c, axion decays to Higgses. Ratio of matter radiation equality scale factor $a_{eq}$ over the one from $\Lambda$CDM vs. the ratio between the axion and the inflaton mass. Underneath the pink line, where the ratio is $1$, the produced heavy axions are ``overclosing" the universe. The parameters used for computing this plot are corresponding to \cref{table}.} 
	\label{fig:DnEff}
\end{figure}

There are two possible constructions of the SM compatible with fibre inflation. These correspond to Cases I and II discussed earlier. In Case I, the SM resides on a stack of D7-branes that wrap the same 4-cycle as the inflaton. This configuration leads to an unsequestered SM, where the soft terms are of the same order as the gravitino mass. Since the inflaton instead is much lighter than the gravitino, its decay into supersymmetric particles is kinematically forbidden~\cite{Cicoli:2018cgu}. Consequently, the primary decay channels of the inflaton are into Higgs bosons, SM gauge bosons, and hidden sector axions.
In this case, the branching ratio of inflaton perturbative decay into the two axions is small, significantly less than unity. Parametric resonance effects will non-perturbatively produce light axions, contributing to dark radiation, but the observational effect will remain in agreement with observational constraints ($\Delta N_\eff \lesssim 10^{-6}$). 

If, however, the produced axions are heavier, their interactions with the SM particles will cause them to decay into visible sector particles. This would accelerate the reheating process, leading to a higher reheating temperature.

In Case II, the SM is living on D3-branes at singularities, and it is sequestered from the bulk, resulting in an effective decoupling of the inflaton sector from the visible degrees of freedom. This case in its most simple realization has been ruled out in~\cite{Cicoli:2018cgu} because it produces too much dark radiation. A later construction~\cite{Cicoli:2022uqa} solves this problem by considering a more general moduli-dependence of the Giudice-Masiero term that allows to considerably reduce the production of dark radiation by introducing an additional coupling between the inflaton fields and the Higgses in the K\"ahler potential (cf. \cref{GiudiceMasiero}). 

In Case II-a, if the produced axions are heavy and remain stable, they contribute to the universe’s dark matter energy density. In this case, if their density is too high, they could lead to the overclosure of the universe, thus imposing stringent constraints on the model parameters to avoid such a scenario.
The heavier the axion, the earlier the time of matter-radiation equality is reached. We can use \cref{matdom} to find the maximum value of the axion mass such that we do not obtain an early matter radiation equality: we want $\alpha_0=\frac{a_{eq}}{a_{eq}^{\Lambda \text{CDM}}}<1$. This provides an upper bound on $m_\vartheta$, depending on the model parameters. In \cref{fig:DnEff}, we plot this ratio as the blue line. Since the produced axions remain as non-relativistic particles and do not decay, values below the pink line in the shaded blue region ($\alpha_0=1$) indicate overproduction of dark matter, leading to an overclosed universe. Therefore this leads to a bound on the axion mass $m_\vartheta\lesssim 10^{-19}\minf\sim 10^{-2}$ MeV. 
However, we stress that this is an idealized benchmark case, as in a realistic scenario the heavy fields will find some channel through which they decay.

Case II-b describes the case in which the axions decay products are massive. Here, we need to distinguish further between two cases: if they decay into massive non-relativistic fields or into massive but relativistic fields. In both cases, the universe still faces the risk of overclosure. In the first case we are in a similar situation as in the stable axion scenario II-a, and the expression for $a_{eq}/a_{eq}^{\Lambda \text{CDM}}$ does not change from \cref{matdom}.
If the decay products are relativistic instead, they will initially redshift as radiation, and as they become non-relativistic, they transition to behaving like matter, modifying the redshift dynamics and slightly relaxing the constraints on the model since the energy density redshifts more rapidly when the particles are relativistic.
The new $a_{eq,1}$ comes from \cref{aeqnew} and we can use \cref{matdomnew} to constraint the axion mass via the new ratio $\alpha_1$. Case II-c instead corresponds to axions decaying into relativistic degrees of freedom, that will therefore contribute to $\Delta N_\eff$ as \cref{dneffdecay}. 

The different possible decay channels are illustrated in the previous section; let us now expand on them. As discussed in \cref{sec:fate}, a light dark sector arising from D7-branes wrapping a blow-up 4-cycle is disfavored. This is because perturbative decays of the inflaton into such a sector tend to overproduce dark radiation. In the context of fiber inflation, the same reasoning applies directly, leading to a similar issue. Consequently, we exclude the possibility of a 7-brane dark sector associated with additional blow-ups in this setup.

We now estimate the decay of the fibre axion in the base axion via kinetic mixing (cf. the left diagram of \cref{fig:feynman_diagr}). Indeed, in fibre inflation, the vev of the fibre modulus defines the vev of the base modulus by keeping the overall volume fixed, and therefore the mass of the base axion is set by fixing the mass of the fibre axion: if one is heavy, the other will be very light. Therefore, the three body decay defined above will be extremely suppressed, as the decay rate $\Gamma_{\vartheta_h\to\vartheta_l\vartheta_l\vartheta_l}$ depends on the mass of the light axion $\vartheta_l$, where now $\vartheta_l=\vartheta_2$ and $\vartheta_h=\vartheta_1$. 
For the values we chose in \cref{table}, $m_{\vartheta_1}\simeq 8.6 \times 10^{-5}\mpl$, and $m_{\vartheta_2}\sim 10^{-370}\mpl\sim 0$.

The two other interactions in \cref{fig:feynman_diagr} do not require any direct coupling, but rely on gravitational interaction. 
First, the axion can decay to gravitons via a two body decay with decay rate 
\begin{equation}
    \Gamma_{\vartheta\to h h} = \frac{|g_{\vartheta    hh}|^2 m_\vartheta^7}{\mpl^4 512 \pi}\,,
\end{equation}
where $g_{\vartheta    hh}\sim \frac{1}{384\pi^2 f_{\vartheta}}$ is derived in \cref{ahhcoupling}.  
The time for the axion to decay into the gravitons therefore reads
\begin{equation}
a_{dec}\sim \frac{1}{T_{dec}}\simeq \left(\frac{g_* 8\pi^3}{90}\right)^{1/4}\frac{1}{\Gamma^{1/2}\mpl^{1/2}}=\left(\frac{g_* 8\pi^3}{90}\right)^{1/4}\frac{512^{1/2} \pi^{1/2} }{m_\vartheta^{7/2} |g_{\vartheta    hh}| \mpl^{-3/2}}\,,
\end{equation}
contributing to $\Delta N_\eff$ as 
\begin{equation}
\begin{aligned}
    \Delta N_\eff&=\frac{120}{7\pi^2}\left(\frac{11}{4}\right)^{\frac{4}{3}}\frac{\rho_\vartheta^{max}}{T_{reh}^4}\left(\frac{a_{max}}{a_{end}}\right)^4\left(\frac{a_{end}}{a_{reh}}\right)^4\left(\frac{g_* 8\pi^{3}}{90}\right)^{\frac{1}{4}}\left( \frac{512 \pi \mpl^3}{m_\vartheta^{5} |g_{\vartheta    hh}|^2 }\right)^{\frac{1}{2}}\left(\frac{s_{2}}{s_{1}}\right)^{\frac{1}{3}}  \,,
\end{aligned}
\end{equation}

Another interesting possibility is the scattering $\vartheta\vartheta\to xx$, where with $xx$ we mean both hidden sector and visible sector particles, as it is a gravitational interaction. 
The decay rate, independently of the decay products, can be approximated, as explained in the above section, by
\begin{equation}
        \Gamma_{\vartheta\vartheta\to xx}\sim \frac{1}{1024 \pi }\frac{\minf^3}{ \mpl^2}\,.
\end{equation}
This interaction will lead to a contribution to $\Delta N_\eff $ when $x$ is a dark sector or the light axion --- even if a negligible one --- while it will contribute and fasten the reheating process when $x$ is a SM particle. 
There will therefore be a branching ratio that takes into account the amount of visible or hidden degrees of freedom if we want to compute these contributions in any given sufficiently concrete specific string model with an explicit SM-like sector. 

Finally, we consider the terms in \cref{GiudiceMasiero} used to obtain a direct channel with the MSSM and achieve an efficient reheating. These introduce additional derivative couplings between the axion and the Higgs fields, that lead to a scattering $\vartheta \vartheta\to HH$, as we derive in \cref{app:higgsdecay}. 
The decay rate (considering $y_1=1$ and $y_2$=0) reads 
\begin{equation}
\Gamma_{\vartheta\vartheta\to HH}\sim \frac{1}{64\pi}\frac{\rho_\vartheta}{m_\vartheta}\frac{1}{m_\vartheta^2+|\vec{p}_\vartheta|^2}\frac{f_\vartheta^2 |\vec{p}_\vartheta|^4}{\mpl^6}\,.
\end{equation}
Since the resonance is most efficient when the momentum of the produced axion is $|\vec{p}_\vartheta|\sim \minf/2$, we can estimate
\begin{equation}
\Gamma_{\vartheta\vartheta\to HH}\sim \frac{1}{4 \times 64\pi}\frac{\rho_\vartheta}{m_\vartheta}\frac{1}{1+\frac{4m_\vartheta^2}{\minf^2}}\frac{f_\vartheta^2 \minf^2}{\mpl^6}\,.
\end{equation}
This decay will contribute to the reheating of the standard model, if efficient enough. However, before decaying, this axion behaves as matter, and will contribute to dark matter. Depending on its mass and its energy density, it might take over the energy budget of the universe before $a_{eq}^{\Lambda \text{CDM}}$, and therefore overclose the universe. 
To check this, we need to estimate $\alpha_1\equiv \frac{a_{eq,1}}{a_{\Lambda \text{CDM}}}$. We can estimate the time of decay as:
\begin{equation}
   T_{dec}=\left(\frac{90}{8\pi^3 g_*}\right)^{1/4}  \Gamma^{1/2}\mpl^{1/2}\,.
\end{equation}
Therefore, 
\begin{equation}
\begin{aligned}
a_{dec}\,\sim\, \frac{1}{T_{dec}}&\,\simeq \left(\frac{g_* 8\pi^3}{90}\right)^{1/4}\frac{1}{\Gamma^{1/2}\mpl^{1/2}}\\
&=\left(\frac{g_* 256^2 8\pi^5}{90}\right)^{1/4} \left( \frac{m_\vartheta}{\rho_\vartheta}(1+\frac{4m_\vartheta^2}{\minf^2}) \right)^{1/2} \frac{\mpl^{5/2}}{f_\vartheta \minf}\,.
\end{aligned}
\end{equation}
With \cref{matdomnew}, we find
\begin{equation}
\begin{aligned}
    \alpha_1&=\,\,\alpha_0\frac{a_{NR}}{a_{dec}}=\,\,\alpha_0\frac{T_{dec}}{m_H}=\\
    &=\,\,\alpha_0 \left(\frac{90}{g_* 256^2 8\pi^5}\right)^{1/4}\left(1+\frac{4m_\vartheta^2}{\minf^2}\right)^{- 1/2}\frac{f_\vartheta \minf}{\mpl^2}\frac{\rho_\vartheta^{1/2}}{m_\vartheta^{1/2}\mpl^{1/2}m_H}\,,
\end{aligned}
\end{equation}
where we defined $\alpha_0$ in \cref{matdom} as the ratio between the matter radiation equality in the $\Lambda $CDM model and in the idealized case where the axion does not decay. We plot the two curves $\alpha_0$ and $\alpha_1$ in \cref{fig:DnEff}. This case is much less constraining, as we expected, with respect to the idealized case where the axion did not decay.
The axion mass is now constrained to $m_\vartheta\lesssim 10^{-8}\minf\sim 100\,$TeV.
%


\subsection{Influence of axion initial conditions}

Our approach relies on taking the initial condition of the axion for parametric resonance to be $\langle\vartheta\rangle =0$. However, this may not be the state of the axion field after inflation. During inflation, the axion can undergo a random walk if its mass is below the Hubble scale $H_\infl$ during inflation. If the axion walks, the initial condition for parametric resonance may be non-zero. As long as the displacement of the axion at the end of inflation is such that the effective axion mass is positive, one can modify the discussion above by keeping track of the axion zero mode. However, if $m_{\vartheta,\eff}^2<0$ initially, there will be a tachyonic instability which will lead to a non-perturbative production of axions as well --- we do not enter into details as this was already well studied in~\cite{Kofman:2001rb}. This issue is relevant only if the axion is sufficiently lighter than $H$. Above we have illustrated that parametric resonance for closed string axions is most efficient when the axion mass is roughly that of the inflaton. In this case, the axion will not undergo large displacements from random walks during inflation and instead finds its minimum rapidly.


\section{Conclusions}
\label{sec:conclusions}

Certain string compactifications give rise to axion potentials that depend on an exponential function of the K\"{a}hler moduli. Identifying one of these moduli with the inflaton results in an axion mass that varies with time as the inflaton oscillates about the minimum of its potential, which sets the stage for parametric resonance.  However, this exponential coupling gives rise to different phenomenology compared to typical parametric resonance studies. Furthermore, the inflaton couples kinetically with the axion, and so the axion equation of motion is described by the Whittaker-Hill equation, a generalization of the typical Mathieu equation found in preheating literature. 

In this framework, particle production occurs most effectively when the inflaton reaches its maximum negative displacement relative to its vacuum expectation value, where the produced particles remain light. As the inflaton oscillates to positive values, these particles become heavy. In an expanding universe, the oscillations of the inflaton field are dampened due to the Hubble expansion. This damping plays a key role because it gradually shifts the resonance from being broad, where particle production is very efficient, to narrow, where particle production becomes much less efficient, before eventually slowing it to a halt. In the case of string inflation, the damping is even more severe. The oscillations in the exponential coupling lead to exponential damping, which means that the conditions for resonance can change dramatically compared to standard QFT models. This drastic damping highlights an important point: it is not enough to simply compare the growth rate of resonance in a non-expanding universe to the expansion rate of the universe in order to predict resonance in an expanding universe. In fact, the exponential suppression due to damping must be carefully accounted for to understand when and if resonance occurs.

Moreover, the oscillation strength, and consequently the particle production rate, depends on the axion mass, as this is determined by the same coupling that drives the oscillations. Parametric resonance produces a very light axions primarily through kinetic mixing, independent of the non-perturbative mass terms. In contrast, for a heavier axion, the instanton contribution to the superpotential dominates production via parametric resonance. 
This difference arises because the kinetic mixing produces effectively a bilinear coupling between inflaton and axion resulting in small amounts of production by parametric resonance. The coupling provided by the instanton, however, is highly non-linear and thus drives strong parametric resonance unless its overall scale is suppressed by dialing the axion mass to be small. 

This results in two major classes of outcomes. 
Case I characterizes axions that are light enough to contribute to the effective number of relativistic degrees of freedom $\Delta N_\eff$. On the other hand, Case II describes the situation where the axion is heavy. From the time of its production onward, it can either stay heavy throughout the cosmic evolution (which we call Case II-a), or it can decay into massive stable fields (Case II-b), or lastly it can decay into massless or very light particles (Case II-c). 
In Cases II-a and II-b, the axion makes up a fraction of DM, depending on how much and how energetically they are produced. Bounding the amount of DM today with $\Omega_{DM}h^2\simeq 0.12$ we can put a bound on the axion mass.
Case II-c instead characterizes a universe filled with non-relativistic degrees of freedom --- the decay products of the axions. These contribute to dark matter abundance, but the details depend on the decay rate, the mass of the final particles, and other details of the compactification and the inflationary model under consideration. We provide expressions for each Case, including final observable quantities and the associated limits. As an illustrative example, we examine fibre inflation, considering the evolution of both the inflaton and axion fields.

In order to simplify the study of parametric resonance, we considered scenarios where complicating factors such as self-resonance of the inflaton field, or tachyonic preheating from an imaginary effective axion mass, could be safely neglected. We achieved this by choosing to assign as initial condition of the classical mode of the axion field the minimum of the potential, such that $m_{\vartheta,\eff}^2>0$. 
However, in models of inflation where $H>m_\vartheta$, the axion field value can undergo random walks. In the case where the axion field initially is displaced from its vev, one needs to take into account also the motion of the axion field towards the minimum of the potential. This can lead to a delay in the start of preheating, or to tachyonic reheating if $m_{\vartheta,\eff}^2<0$. We note that parametric resonance in string inflation may also allow for a process akin to ``instant reheating," where particles decay almost immediately after being produced, in the case where the axion decay is very rapid and is energetically allowed.

Finally, we wish to highlight that for fibre inflation there is no self resonance of the inflaton quanta, as was found instead for blow-up inflaton in~\cite{Barnaby:2009wr}. 
The difference lies in the very sharp minimum of the blow-up modulus inflationary potential, such that~\cite{Barnaby:2009wr} finds a periodic tachyonic instability where the effective mass of the inflaton field becomes imaginary. 
The scalar potential that characterizes fibre inflation does not have a sharp minimum, and therefore also the oscillations of the inflaton field around its minimum will not be that violent. The effective mass therefore can, but does not have to, become imaginary. 
The effect of parametric resonance as a non-perturbative mechanism to produce axions from preheating is most visible when it is the lone non-perturbative effect present, and therefore we assume to be in a situation where we can neglect these additional phenomena. 
We note that for a complete analysis, one should consider both of these types of preheating together, on a lattice.
For our purposes, that is, to illustrate the role parametric resonance plays in a non-perturbative production of axions, staying below the limit of inflaton fragmentation and backreaction, separating the two effects is a reasonable assumption.

We conclude by considering the observational implications of our analysis. Parametric resonance provides a means to produce a Cosmic axion Background (CaB) of ultralight axions utilizing the kinetic coupling of the inflaton to axions. However, this population is small and lies far below upcoming experimental probes of $\Delta N_{\eff}$. Nonetheless, these CaB populations are unavoidable and may be observable in future experiments. On the other hand, heavy axions are much more efficiently produced. Thus parametric resonance in string inflation could be utilized as a means to produce dark matter or alternative routes of reheating. These results were largely couched in the framework of type IIB string theory, and it would be interesting to understand the situation in other perturbative frameworks, such as heterotic compactifications. Beyond these considerations, parametric resonance can produce compact objects such as oscillons or axion stars. This would require a lattice treatment, but would nonethless be a natural extension of the current work in order to completely categorize the observational consequences of parametric resonance of axions in string inflation. Finally, we expect parametric resonance to produce spectator axions via kinetic or instanton couplings also in setups where inflation is driven by a string axion as well. We leave this interesting issue for future work.

 
\acknowledgments
We thank L.\,Brunelli, L.\,Caraffi, M.\,Cicoli, D.\,J.\,E.\,Marsh and F.\,G.\,Pedro for useful discussions. JML, MP, and NR 
gratefully acknowledge support from the Simons Center for Geometry and Physics, Stony Brook University for its hospitality during the completion of a portion of this work. NR is supported by a Leverhulme Trust Research Project Grant RPG-2021-423. This article is based in part upon work from COST Action COSMIC WISPers CA21106, supported by COST (European Cooperation in Science and Technology). MP is supported by the Deutsche Forschungsgemeinschaft under Germany’s Excellence Strategy - EXC 2121 “Quantum Universe” - 390833306. AW is partially supported by the Deutsche Forschungsgemeinschaft under Germany’s Excellence Strategy - EXC 2121 “Quantum Universe” - 390833306 and by the Deutsche Forschungsgemeinschaft through the Collaborative Research Center SFB1624 ``Higher Structures, Moduli Spaces, and Integrability’'. This work was co-funded by the European Union and supported by the Czech Ministry of
Education, Youth and Sports (Project No. FORTE – CZ.02.01.01/00/22\_008/0004632).

\appendix
\section{Gravitational Chern-Simons Couplings in Type IIB Orientifolds}
\label{app:gravitationalCS}
In this appendix we derive the coupling of 4D $C_4$ axions to $R\widetilde{R}$ terms in Type IIB orientifolds. Following~\cite{Jockers:2004yj}, we assume a product ansatz for spacetime of the form $\mathbb{R}^{3,1}\times \widetilde{X}_6$. Here $\widetilde{X}_6 = X_6/\mathcal{P}$ is the orientifold, while $X_6$ is a Calabi-Yau 3-fold and $\mathcal{P}$ is an orientifold projection defined via an isomorphic and holomorphic involution $\sigma$ of $X_6$. We assume a product metric:
\begin{equation}
   g_{10} := g_{\mu\nu}\;dx^\mu\otimes dx^\nu + 2g_{i\bar{j}}(y)\; dy^i \otimes d\bar{y}^{\bar{j}}\,.
\label{eq:metansatz}
\end{equation}
The effective 4D action bulk terms can be obtained 
from dimensional reduction of the 10D type IIB fields. In particular, the $C_4$ expansion is
\begin{equation}
    C_4 = \rho_\alpha(x) \widetilde{\omega}^\alpha(y)+\dots\,,
\end{equation}
where the $\{\widetilde{\omega}^\alpha\}$, $\alpha = 1,..,h^+_{2,2}$, form a basis for $H^{(2,2)}_{\bar{\partial},+}(X_6)$
We will assume that there is a D7-brane permeating spacetime and wrapping a 4-cycle $S_+$ of $\widetilde{X}_6$ such that the worldvolume of the brane is
\begin{equation}
    \mathcal{W} := \mathbb{R}^{3,1}\times S_+\,.
\end{equation}
We will denote the embedding map of the worldvolume as $\varphi: \mathcal{W} \xhookrightarrow{} \mathbb{R}^{3,1}\times \widetilde{X}_6$. Note also that $S_+\in H_4(X_6,\mathbb{Z})$ and is the union of two 4-cycles $S_1$ and $S_2$ in $X_6$. We will denote the embedding map as $\iota: S_+ \xhookrightarrow{} X_6$. 
The pullback of~\cref{eq:metansatz} is
\begin{equation}
    \varphi^*g_{10} = g_{\mu\nu} dx^\mu \otimes dx^\nu + 2g_{i\bar{j}}(y)\; dy^i \otimes d\bar{y}^{\bar{j}} + 2 g_{i\bar{j}}(y) \;\partial_\mu \xi^i \partial_\nu \bar{\xi}^{\bar{j}} dx^\mu \otimes dx^\nu\,,
\end{equation}
where the internal part is suitably restricted to the wrapped 4-cycle. For the moment, we will set the fluctuations $\xi^i =0$. Then the worldvolume metric can be written as 
\begin{equation}
    \varphi^* g_{10} =  \eta_{\alpha\beta}\theta^\alpha \otimes \theta^\alpha + 2\delta_{I\bar{J}} \widetilde{\theta}^I\wedge \widetilde{\theta}^J\,,
\end{equation}
where $\theta^\alpha := e^\alpha_{\;\; \mu }dx^\mu$ and $\widetilde{\theta}^I := e^I_{\;\; i} dy^i$ define a non-coordinate basis via the external and internal vielbeins. Since the internal and external vielbeins depend only on external and internal coordinates, respectively, the Cartan structure equations imply that the connection 1-form splits into independent external and internal pieces. The curvature 2-form follows the same behavior. Therefore, the curvature 2-form of the tangent bundle is block-diagonal between the external and internal pieces.   

The relevant parts of the D7-brane Chern-Simons action are
\begin{equation}
    S_{D7}\supset \mu_7 \int_{\mathcal{W}} \frac12 (2\pi\alpha^\prime)^2 \varphi^*(C_4) \wedge \frac{N}{48}\Big(\text{tr}(R_T \wedge R_T) - \text{tr}(R_N\wedge R_N)\Big)\,.
\end{equation}
Where $R_{T/N}$ are the curvature forms of the tangent/normal bundle of $\mathcal{W}$. From the above, we know that
\begin{equation}
   \text{tr} (R_T\wedge R_T) = \text{tr} (R_{ST}\wedge R_{ST}) + \text{tr} (R_{+}\wedge R_+)\,.
\end{equation}
Thus we find a 4D coupling
\begin{equation}
    \begin{aligned}
        S_{4D}&\supset \frac{N\mu_7}{48} \int_{M_4} \frac12 (2\pi\alphap)^2\ell_S^4\; \rho_+\; \text{tr} (R_{ST}\wedge R_{ST})\\
        &= \frac{N}{192\pi}\int_{M_4}\rho_+ \text{tr}(R_{ST}\wedge R_{ST})\\
        &= \frac{2\pi N}{48} \int \rho_+ p_1(R_{ST}) \,.
    \end{aligned}
\end{equation}
Where $p_1(R_{ST})$ is the first Pontrjagin class. For a 4D spin manifold, the integral of Pontrjagin class is quantized as
\begin{equation}
    \int_{X_4} p_1(R_2) = 48m
\end{equation}
with $m\in\mathbb{Z}$. Thus $e^{iS_{4D}}$ is invariant under $\rho_+\rightarrow \rho_++n$ with $n\in\mathbb{Z}$. We can define a canonically-normalized axion via
\begin{equation}
    \rho_+  =\frac{\vartheta}{2\pi f_\vartheta}
\end{equation}
Then we have a coupling
\begin{equation}
   S_{4D} \supset \frac{N}{1536\pi^2}\int_{M_4} \frac{\vartheta}{f_{\vartheta}} R^{\mu\nu}_{\;\;\; \rho\sigma} R_{\mu\nu \alpha\beta}\varepsilon^{\rho\sigma\alpha\beta}  d^4x
\end{equation}
We can now define the coupling $g_{\theta  hh}$ we use in the action:
\begin{equation}
    \mathcal{L}\supset \frac{g_{\theta  hh}}{4}\vartheta \epsilon_{\mu\nu\rho\sigma}R^{\mu\nu}_{\;\;\; \alpha\beta}R^{\rho\sigma\alpha\beta}\,,
\end{equation}
\begin{equation}
\label{ahhcoupling}
    g_{\theta  hh}=\frac{N}{384\pi^2 f_\vartheta}\,.
\end{equation}

\section{Axion Higgs coupling}
\label{app:higgsdecay}

In this appendix we derive the contribution to the kinetic terms of the axion fields coming from the Giudice-Masiero terms in the K\"ahler potential (cf. \cref{GiudiceMasiero}), introduced in order to allow a reheating process. 

Recall that the kinetic terms will be given by $K_{i\bar{j}}\partial_\mu \theta^i\partial^\mu\theta^{\bar{j}}$ where the K\"ahler metric is given by 
\begin{equation}
K_{i\bar{j}}=\partial_{T_i}\partial_{\bar{T_j}}K \,.
\end{equation}
Let us then consider the Giudice-Masiero terms of the form: 
\begin{equation}
    K\supset \frac{H\bar{H}}{\left(T_1+\bar{T}_1\right)^{y_1}\left(T_2+\bar{T}_2\right)^{y_2}}\,.
\end{equation}
The contribution to the kinetic terms then reads: 
\begin{equation}
\begin{aligned}
    \mathcal{L}\supset& \frac{y_1(y_1+1) }{\left(T_1+\bar{T}_1\right)^{y_1+2}\left(T_2+\bar{T}_2\right)^{y_2}} H\bar{H}\partial_\mu \theta_1\partial^\mu\theta_1\\
    &+\frac{y_1 y_2 }{\left(T_1+\bar{T}_1\right)^{y_1+1}\left(T_2+\bar{T}_2\right)^{y_2+1}} H\bar{H}\partial_\mu \theta_1\partial^\mu\theta_2\\
    &+\frac{y_2(y_2+1) }{\left(T_1+\bar{T}_1\right)^{y_1}\left(T_2+\bar{T}_2\right)^{y_2+2}} H\bar{H}\partial_\mu \theta_2\partial^\mu\theta_2\,.
\end{aligned}
\end{equation} 
Canonically normalizing the axions with the decay constants derived from the tree level K\"ahler potential $f_i\sim \frac{1}{\left(T_i+\bar{T_i}\right)}$ in units of $\mpl$, we can rewrite the equation above as 
\begin{equation}
\begin{aligned}
    \mathcal{L}\supset& H\bar{H}f_1^{y_1}f_2^{y_2} \left(y_1(y_1+1) \partial_\mu \vartheta_1\partial^\mu\vartheta_1+y_1 y_2\partial_\mu \vartheta_1\partial^\mu\vartheta_2+y_2(y_2+1) \partial_\mu \vartheta_2\partial^\mu\vartheta_2\right)\,.
\end{aligned}
\end{equation} 
If we now reduce to the simple case where $y_1=1$ and $y_2=0$, the above reduces to the single term 
\begin{equation}    \mathcal{L}\supset 2f_1 H\bar{H}  \partial_\mu \vartheta_1\partial^\mu\vartheta_1\,.
\end{equation}
This corresponds to a $2\to2$ scattering for which we can simply compute the cross-section and the corresponding decay rate, knowing now that the vertex is $\sim f_1 |p_\vartheta|^2$ in units of $\mpl$. The differential cross section thus reads
\begin{equation}
    \frac{d\sigma}{d\Omega}_{CM} = \frac{1}{64\pi^2 s}|\mathcal{M}|^2 \frac{|p_f|}{|p_i|}\,,
\end{equation}
where the Feynman amplitude $|\mathcal{M}|\sim \left|\frac{f_q p_\vartheta^2}{\mpl}\right|$ and $s=(p+p')^2=4(m_\vartheta^2+|p_\vartheta|^2)$. 
The scattering rate $\Gamma_{\theta\theta\to HH} = n\sigma v$ therefore reads: 
\begin{equation}
\Gamma_{\vartheta\vartheta\to HH}\sim \frac{1}{64\pi}\frac{\rho_\vartheta}{m_\vartheta}\frac{1}{m_\vartheta^2+|\vec{p}_\vartheta|^2}f_\vartheta^2 |\vec{p}_\vartheta|^4 \frac{|\vec{p_\vartheta}|}{m_\vartheta}\sqrt{1+\frac{m_\vartheta^2}{m_\varphi^2}}\,.
\end{equation}
In the above we used the fact that $n_\vartheta=\rho_\vartheta /m_\vartheta$ and $v=\frac{p_\vartheta}{m_\vartheta}$.


\bibliographystyle{JHEP}
\bibliography{CaBiverseRefs}
\end{document}